\documentclass[a4paper,12pt]{article}
\pdfoutput=1
\usepackage{geometry}
\usepackage{amsmath,amssymb,amsfonts}
\usepackage{xcolor,graphicx,cite,soul,ulem}
\usepackage{array,booktabs,longtable}
\usepackage{caption,microtype}
\usepackage{hyperref}

\definecolor{darkblue}{rgb}{0,0,.75}
\definecolor{darkgreen}{rgb}{0,.75,0}
\definecolor{darkred}{rgb}{.75,0,0}
\hypersetup{colorlinks=true,
  linkcolor=darkred,
  citecolor=darkgreen,
  urlcolor=darkblue}

\newcommand{\lsim}
{\;\raisebox{-.3em}{$\stackrel{\displaystyle <}{\sim}$}\;}
\newcommand{\gsim}
{\;\raisebox{-.3em}{$\stackrel{\displaystyle >}{\sim}$}\;}

\newcommand\Code[1]{\ensuremath{\texttt{#1}}}

\newcommand\al{\alpha}

\newcommand\tb{\tan\beta}
\newcommand\TB{t_\beta}
\newcommand\SB{s_\beta}
\newcommand\CB{c_\beta}

\newcommand\SBA{s_{\beta - \alpha}}

\newcommand\LP{\left(}
\newcommand\RP{\right)}
\newcommand\LB{\left[}
\newcommand\RB{\right]}
\newcommand\LV{\left\{}
\newcommand\RV{\right\}}

\newcommand\ReTilde{\mathop{\widetilde{\mathrm{Re}}}}
\newcommand\ReDiag{\mathop{%
  \raise .5pt\hbox{[}%
  \widetilde{\mathrm{Re}}%
  \raise .5pt\hbox{]}}}
\newcommand\ReOffDiag{\mathop{%
  \raise .5pt\hbox{$\llbracket$}%
  \widetilde{\mathrm{Re}}%
  \raise .5pt\hbox{$\rrbracket$}}}
\newcommand\Sig{\Sigma}

\newcommand\OS{\ensuremath{\mathrm{OS}}}
\newcommand\DRbar{\ensuremath{\smash{\overline{\mathrm{DR}}}}}
\newcommand\MSbar{\ensuremath{\overline{\mathrm{MS}}}}

\newcommand\SW{s_\mathrm{w}}
\newcommand\CW{c_\mathrm{w}}
\newcommand\MW{M_W}
\newcommand\MZ{M_Z}
\newcommand\Mh{M_h}
\newcommand\MH{M_H}
\newcommand\MA{M_A}
\newcommand\MHp{M_{H^\pm}}
\newcommand{\hotf}{\ensuremath{h_{125}}}

\newcommand\mt{m_t}

\newcommand\At{A_t}

\newcommand\Sf{{\tilde f}}

\newcommand\dTB{\delta\TB}
\newcommand\ino[1]{\tilde\chi_{#1}}

\newcommand\chapm[1]{\ino{#1}^\pm}

\newcommand\cha{\chapm}
\newcommand\mcha[1]{m_{\chapm{#1}}}

\newcommand\neu[1]{\ino{#1}^0}
\newcommand\mneu[1]{m_{\neu{#1}}}

\newcommand\refeq[1]{Eq.~(\ref{#1})}

\newcommand\refta[1]{Tab.~\ref{#1}}
\newcommand\refse[1]{Sect.~\ref{#1}}

\newcommand\citere[1]{Ref.~\cite{#1}}
\newcommand\citeres[1]{Refs.~\cite{#1}}

\newcommand\eg{e.g.\ }
\newcommand\ie{i.e.\ }

\newcommand{\CP}{{\cal CP}}

\newcommand{\onel}{one-loop}

\newcommand{\gev}{\,\, \mathrm{GeV}}
\newcommand{\mev}{\,\, \mathrm{MeV}}

\newcommand\FA{\texttt{FeynArts}}
\newcommand\FC{\texttt{FormCalc}}
\newcommand\LT{\texttt{LoopTools}}
\newcommand\FH{\texttt{FeynHiggs}}

\newcommand\HDECAY{\texttt{HDECAY}}

\newcommand\fb{\ensuremath{\mbox{fb}}}

\newcommand\mh[1]{m_{h_{#1}}}

\newcommand\mstop[1]{m_{\tilde{t}_{#1}}}

\def\order#1{\ensuremath{{\cal O}(#1)}}
\def\reffi#1{\mbox{Fig.~\ref{#1}}}
\def\reffis#1{\mbox{Figs.~\ref{#1}}}
\def\als{\alpha_s}

\def\ga{\gamma}
\def\de{\delta}
\def\ka{\kappa}
\def\la{\lambda}
\def\mueff{\mu_{\mathrm{eff}}}

\def\phiAt{\varphi_{\At}}

\definecolor{Orange}{named}{orange}
\definecolor{Purple}{named}{purple}
\definecolor{Lightblue}{cmyk}{0.9,0.1,0.1,0.3}
\definecolor{dgelborange}{cmyk}{0.,0.3,0.5, 0.}
\definecolor{Lila}{rgb}{0.5,0.,1}

\graphicspath{{figs/}}
\allowdisplaybreaks
\begin{document}
\thispagestyle{empty}

\def\thefootnote{\fnsymbol{footnote}}

\begin{flushright}
\mbox{}
IFT--UAM/CSIC-25-065 %\\
%arXiv:\htr{XXYY.ZZZZZ} [hep-ph]
\end{flushright}

\vspace{0.5cm}

\begin{center}

{\large\sc 
{\bf Light Neutral Higgs-Boson Production at \boldmath{$e^+e^-$} Colliders}} 

\vspace{0.4cm}

{\large\sc {\bf in the Complex MSSM and NMSSM: A Full One-Loop Analysis}}

\vspace{1cm}

{\sc 
S.~Heinemeyer$^{1}$%
\footnote{email: Sven.Heinemeyer@cern.ch}%
, S.~Passehr$^{2}$%
\footnote{email: sebastian.passehr@web.de}\footnotemark[4]%
~and C.~Schappacher$^{3}$%
\footnote{email: schappacher@kabelbw.de}\footnotemark[4]%
\footnotetext[4]{former address}% 
}

\vspace*{.7cm}

{\sl
$^1$Instituto de F\'isica Te\'orica (UAM/CSIC), Universidad
  Aut\'onoma de Madrid,\\ Cantoblanco, 28049, Madrid, Spain

\vspace*{0.1cm}

$^2$Institute for Theoretical Particle Physics and Cosmology,\\
RWTH Aachen University, Sommerfeldstra\ss e 16, 52074 Aachen, Germany

\vspace*{0.1cm}

$^3$Institut f\"ur Theoretische Physik,
Karlsruhe Institute of Technology, \\
76128 Karlsruhe, Germany
}

\end{center}

\vspace*{0.1cm}

\begin{abstract}
\noindent
For future precision analyses of the Higgs boson at $\approx 125\gev$,
\hotf, a precise knowledge of its production and decay properties is
mandatory. While in the Standard Model~(SM) these calculations are
quite advanced, in many models beyond the SM~(BSM) a precise
calculation is missing so far. We present the calculation of the
Higgs-strahlung cross-sections at $e^+e^-$\,colliders for the light
neutral Higgs boson production in the Next-to-Minimal Supersymmetric
SM~(NMSSM) with complex parameters~(cNMSSM).  The evaluation is based
on a full one-loop calculation of the production mechanism
$e^+e^- \to h_1 Z$, including soft, hard, and collinear photon radiation.
The dependence of the Higgs boson production cross-sections on the
relevant cNMSSM~parameters is analyzed numerically.
In certain scenarios we find sizable corrections
to the Higgs-strahlung cross-section. Normally, they reach
about~$10\%$ of the tree-level results, but can also exceed 20\%.
Finally, the calculation is compared to the
corresponding one in the Minimal Supersymmetric SM~(MSSM).
The knowledge of the full one-loop contributions
to the Higgs-boson production is particularly important for
a sound theoretical interpretation of
measurements at future $e^+e^-$\,colliders such as
the~ILC, CLIC, LCF, FCC-ee, or~CEPC.  It is planned to implement the
evaluation of the Higgs boson production cross-sections into an add-on
package to the code \FH. 
\end{abstract}

%\pacs{}

\def\thefootnote{\arabic{footnote}}
\setcounter{page}{0}
\setcounter{footnote}{0}

\newpage

%%%%%%%%%%%%%%%%%%%%%%%%%%%%%%%%%%%%%%%%%%%%%%%%%%%%%%%%%%%%%%%%%%%%%%%%%%%%%%%
%%%%%%%%%%%%%%%%%%%%%%%%%%%%%%%%%%%%%%%%%%%%%%%%%%%%%%%%%%%%%%%%%%%%%%%%%%%%%%%

\section{Introduction}
\label{sec:intro}

The discovery of a new particle with a mass of about $125\gev$ in the
Higgs searches at the Large Hadron Collider (LHC) about 12~years ago
by ATLAS~\cite{ATLASdiscovery} and CMS~\cite{CMSdiscovery}, marks the
culmination of an effort that has been ongoing for more than half a
century and opens a new era of particle physics.  Within the
experimental and theoretical uncertainties the
measured properties of the newly discovered
particle are, so far, in agreement with a Higgs boson as
predicted in the Standard Model~(SM)~\cite{ATLAS-Higgs,CMS-Higgs}.
However, the experimental results for the state at~$\approx 125\gev$,
whose couplings are known up to now to an experimental precision of
roughly~$\sim 10$-$20\%$, leave ample room for interpretations in
models beyond the SM~(BSM).

The identification of the underlying physics of the discovered new
particle and the exploration of the mechanism of electroweak symmetry
breaking will clearly be a top priority in the future program of
particle physics.
The most frequently studied theories incorporating a Higgs particle with
the measured properties are the SM, and extensions of the SM by another
Higgs doublets and possibly an additional singlet. The latter two
scenarios can be further enriched by incorporating supersymmetry (SUSY)
into the model, yielding the Minimal Supersymmetric SM
(MSSM)~\cite{mssm,HaK85,GuH86} and the Next-to MSSM
(NMSSM)~\cite{Maniatis:2009re,Ellwanger:2009dp}, respectively.
The extension of
the~MSSM by a gauge-singlet superfield was originally motivated by
the~`$\mu$~problem'\,\cite{Kim:1983dt}.
Contrary to the case of the~SM, as mentioned above, two complex Higgs
doublets are required in the~MSSM,
which are extended further by a complex Higgs singlet
in the~NMSSM.  This
results in five or seven physical Higgs bosons in
the~MSSM or~NMSSM instead of the single physical
Higgs boson in the~SM. At lowest order in the~MSSM these are the light
and heavy $\CP$-even Higgs bosons, $h$ and~$H$, the $\CP$-odd Higgs
boson,~$A$, and two charged Higgs bosons,~$H^\pm$.
Considering only real parameters, the~NMSSM extends the $\CP$-even and
$\CP$-odd sectors by one Higgs boson each. In the
case of complex parameters (and taking into account higher-order
corrections for the~MSSM), all neutral Higgs bosons
can mix, resulting in $h_i$ ($i = 1,2,3$) in
the~cMSSM~\cite{mhiggsCPXgen,Demir,mhiggsCPXRG1,mhiggsCPXFD1}, or in
$h_i$ ($i = 1,2,3,4,5$) in the~cNMSSM~\cite{NMSSMCT}.
In this work we will consider in general the~cMSSM and the~cNMSSM.

The Higgs sector of the~cMSSM is described at the tree level by two
parameters: the mass of the charged
Higgs bosons,~$\MHp$, and the ratio of the two vacuum-expectation
values~(\textsc{vev}s), $\tb \equiv \TB = v_2/v_1$.
In addition to the two cMSSM-like parameters, the singlet Higgs
sector of the cNMSSM introduces four additional independent
parameters in the extended Higgs potential. Following the common
nomenclature, we choose:
$\la$, $\ka$, $A_\ka$ and the~\textsc{vev} of the
singlet,~$v_s$.
In the cNMSSM, the mu-Parameter of the cMSSM is generated dynamically
as $\mueff := \la\,v_s$.
In the scenarios investigated in this paper, we
identify the lightest Higgs boson, $h_1$ with the particle discovered
at the LHC at~$\approx 125\gev$~\cite{ATLASdiscovery,CMSdiscovery},
\hotf\ (see \citere{Mh125} for an original analysis).%
\footnote{
    The results of our paper are equally valid for any Higgs boson
    that is identified with the \hotf.
}
Here it should
be noted that in SUSY models the mass of the Higgs bosons other
  than the charged one are not free parameters,
but can be calculated in terms of the other
model parameters, see \citere{Slavich:2020zjv} for a review.  This
sets important constraints on the SUSY parameter
space. 

Several possible future $e^+e^-$\,colliders are currently under
consideration. Circular designs, the FCC-ee~\cite{fcc-ee-web} and the
CEPC~\cite{cepc-web}, are proposed to start with a center-of-mass
energy of~$\sqrt{s} = 250\gev$ (going up to~$\sim 350$--$365\gev$
in the final stage). Linear designs, the ILC~\cite{ilc-web,LCreport}, 
LCF~\cite{LCF} and CLIC~\cite{LCreport,clic-web},
are proposed to start with~$\sqrt{s}
= 250 \gev$ and~$380 \gev$, respectively (going up to~$1000 \gev$
and~$3000 \gev$ in the final stage, respectively).
At a future $e^+e^-$\,collider several production modes for the
neutral Higgs bosons in the~cNMSSM are possible, depending on the
available center-of-mass energy,
\begin{align}
e^+e^- &\to h_i Z,\,
            h_i \ga,\,
            h_i h_j,\,
            h_i \nu \bar\nu,\,
            h_i e^+e^-,\,
            h_i t \bar{t},\, 
            h_i b \bar{b},\,
            \ldots \qquad (i,j = 1,2,3,4,5)\,, %\notag
\end{align}
where we identify $h_1 \equiv \hotf$.  Taking into account the
starting energy of 
the various possible future $e^+e^-$\,colliders, only the production
process
\begin{align}
  e^+e^- &\to \hotf Z
\label{eq:eeh1Z}
\end{align}
is accessible with a high rate%
\footnote{ Also the process $e^+e^-\to \hotf \ga$ is kinematically
accessible, but has a substantially lower
rate~\cite{NeutralHiggsProd,Arco:2020eas,Kanemura:2018esc}.
}%
~at all proposed future experiments. It will serve to measure the
couplings of the \hotf\ with very high accuracy, see
\citere{deBlas:2019rxi} for a review.  Roughly speaking, couplings to
heavier fermions will be determined at about the per~cent level,
whereas the couplings to massive gauge bosons can reach the per~mille
level.  Such a precise determination of the \hotf\ couplings requires
a prediction of its production and decay modes at least at the same
level of accuracy. In the following, we will concentrate on the
corresponding calculations in the~cMSSM and~cNMSSM.

In order to yield a sufficient accuracy, one-loop corrections to the
various Higgs boson production and decay modes have to be considered.
Full one-loop calculations in the~cMSSM for various \hotf\ decays into
SM~fermions, but also into their scalar superpartners
as well as charginos and neutralinos
(which are by now experimentally excluded, except for the decay into a
pair of the lightest neutralinos), have been presented over the last
years~\cite{hff,HiggsDecaySferm,HiggsDecayIno}.  For the decay into
SM~fermions, see also \citeres{hff0,deltab,db2l}.
Decays into SM~gauge~bosons (see also \citere{hVV-WH}) can be
evaluated to a very high precision using the full SM one-loop
result~\cite{prophecy4f} combined with the appropriate effective
couplings~\cite{mhcMSSMlong}.  The full one-loop corrections in
the~cMSSM listed here together with resummed SUSY~corrections have
been implemented into the code
\FH~\cite{feynhiggs,mhiggslong,mhiggsAEC,mhcMSSMlong,Mh-logresum,moreFH}.
Corrections at and beyond the one-loop level in
the~MSSM with real parameters~(rMSSM) are implemented into the code
\HDECAY~\cite{hdecay,hdecay2}.  Both codes were combined by the LHC
Higgs Cross Section Working Group to obtain the most precise
evaluation for~rMSSM Higgs boson decays~\cite{YR3}. 
Corresponding calculations in the cNMSSM can be found in
\citeres{Domingo:2018uim,Domingo:2019vit,Baglio:2019nlc,Dao:2019nxi,Domingo:2020wiy,Dao:2020dfb,Dao:2021jaq,Domingo:2022kfm}. 
These results have not yet been implemented into a public computer code. 

In the above-cited papers it was concluded that the required level of
precision in the \hotf\ decays is not yet reached, but can be done
with known computational techniques. On the other hand, the
corresponding precision level of the \hotf\ production modes is less
advanced.  Taking into account the starting energies of the various
future $e^+e^-$\,collider proposals, from now on we focus on the
process in \refeq{eq:eeh1Z}.  A full one-loop calculation for the process
in \refeq{eq:eeh1Z} in the~cMSSM was presented in
\citere{NeutralHiggsProd} (see also
\citere{Heinemeyer:2016riv}). However, these results have not yet been
included in a public computer code. No corresponding calculation in
the~cNMSSM has been published so far. In this work we present a
consistent calculation of the process in \refeq{eq:eeh1Z} in the~cMSSM
and~cNMSSM at the full one-loop level. The ``MSSM part'' of the
NMSSM~calculation is performed exactly as in the~MSSM. Consequently,
differences in the predictions of the Higgs boson production cross
section can directly be attributed to the differences of the two
models.

The evaluation of the channel (\ref{eq:eeh1Z}) is based on a full
one-loop calculation, \ie including electroweak (EW) corrections, as
well as soft, hard, and collinear photon radiation.
The Higgs boson masses are evaluated at the two-loop
level (using \texttt{NMSSMCALC} (version 5.0)~\cite{nmssmcalc}).
The variety of existing results for the MSSM cross-section
calculations have been summarized in \citere{NeutralHiggsProd}.  On
the other hand, in the~NMSSM very few calculations have been performed
in the past. In \citere{Ham:2007mt} higher-order corrections to the
NMSSM Higgs sector were evaluated in the effective-potential approach
at the one-loop level, taking into account contributions from 3rd
generation quarks and squarks. The corresponding mixing was introduced
into the effective $ZZh_1$~coupling. In \citere{Cao:2014rma} the
production cross-section for a light singlet-dominated Higgs was
evaluated at the tree level. To our knowledge, no full one-loop
calculation in the~cNMSSM has ever been attempted.
Comments about a numerical comparison with the literature will be given in 
\refse{sec:comparisons}.

\medskip

\noindent
In this paper we present for the first time a full and consistent
one-loop calculation for the production of (light, i.e.\
corresponding to \hotf) neutral Higgs bosons in
the~cNMSSM at $e^+e^-$\,colliders in association with a $Z$~boson.  We
take into account soft, hard and collinear photon emission
and the $Z^{\text{mix}}$~factor contributions.  In this way we go
substantially beyond the existing calculations (see above).  In
\refse{sec:renorm} we very briefly review the renormalization of the
relevant sectors of the~cNMSSM.  Details with regards to the
calculation can be found in \refse{sec:calc}, and we also comment on
comparisons with the literature in \refse{sec:comparisons}.  The
numerical results for the channel (\ref{eq:eeh1Z}) are presented in
\refse{sec:numeval}.  The conclusions can be found in
\refse{sec:conclusions}.  There are plans to implement the evaluation
of the production cross-sections into the Fortran code
\FH~\cite{feynhiggs,mhiggslong,mhiggsAEC,mhcMSSMlong,Mh-logresum,moreFH}.

%%%%%%%%%%%%%%%%%%%%%%%%%%%%%%%%%%%%%%%%%%%%%%%%%%%%%%%%%%%%%%%%%%%%%%%%%%%%%%%

\subsection*{Prolegomena}

We use the following short-hands in this paper:
\begin{itemize}

\item $\SW \equiv \sin\theta_{\mathrm w}$, $\CW \equiv \cos\theta_{\mathrm w}$.

\item $\SB \equiv \sin\beta$, 
      $\CB \equiv \cos\beta$, 
      $\TB \equiv \tb$,
      $\SBA \equiv \sin(\beta - \alpha)$.

\end{itemize}
They will be further explained in the text below.

%%%%%%%%%%%%%%%%%%%%%%%%%%%%%%%%%%%%%%%%%%%%%%%%%%%%%%%%%%%%%%%%%%%%%%%%%%%%%%%
%%%%%%%%%%%%%%%%%%%%%%%%%%%%%%%%%%%%%%%%%%%%%%%%%%%%%%%%%%%%%%%%%%%%%%%%%%%%%%%

\section{The complex (N)MSSM}
\label{sec:renorm}

The cross section \refeq{eq:eeh1Z} is calculated at the one-loop level
(including soft, hard and collinear photon radiation), see the next section.
This requires the simultaneous renormalization of the Higgs- and gauge-boson
sectors as well as the electron--positron sector of the c(N)MSSM.
We give a few relevant details as regards these sectors and their
renormalization.  More information can be found in
\citeres{HiggsDecaySferm,HiggsDecayIno,MSSMCT,SbotRen,Stop2decay,%
  Gluinodecay,Stau2decay,LHCxC,LHCxN,LHCxNprod,NMSSMCT}.

The renormalization of the Higgs and gauge-boson sectors of the cMSSM
follows strictly \citere{MSSMCT} and references therein
(see especially \citere{mhcMSSMlong}).  This defines in particular the
counterterm~$\dTB$ as well as the counterterms for the $Z$~boson
mass,~$\de\MZ^2$, and for the sine of the weak mixing angle,~$\de\SW$
(with~$\SW = \sqrt{1 - \CW^2} = \sqrt{1 - \MW^2/\MZ^2}$, where~$\MW$
and~$\MZ$ denote the~$W$ and $Z$~boson masses, respectively).

The renormalization of the fermion sector is described in detail in
\citere{MSSMCT} and references therein.  We use the on-shell (\OS)
renormalization for all three generations of leptons.

Aside from $\de\MZ^2$ and $\de\SW$, and different from the cMSSM, 
also the electric charge $e$ appears as independent parameter in
the tree-level Higgs potential of the cNMSSM and needs to be
renormalized at the one-loop order. We choose to fix it via
\begin{align}
e &\to e\, (1 + \de Z_e)\,, \\
\de Z_{e} &= \frac 12 \LP \frac{\SW}{\CW} \de Z_{Z\gamma} - \de
            Z_{\gamma\gamma}\RP\,, \\
\de Z_{\gamma\gamma} &= -\ReTilde \frac{\partial}{\partial q^2}\Sig_{\gamma}^T(q^2)\Big|_{q^2 = 0}\,, \\
\de Z_{Z\gamma} &= \frac 2{\MZ^2} \ReTilde \Sig_{\gamma Z}^T(0)\,,  
\end{align}
where $\Sig_\ga^T$ and $\Sig_{\ga Z}^T$ denote the transverse
part of the photon and photon-$Z$ self-energies, respectively.
$\ReTilde$ takes the real part of loop integrals only and leaves complex
couplings unaffected, see \citere{MSSMCT}.

In extension to the MSSM,
in the NMSSM an additional \textsc{vev}, $v_S$, is introduced via the
Higgs singlet. The corresponding additional
minimization conditions following from the singlet tadpoles 
are utilized to fix the phases of $\de A_\ka$ and $\de A_\la$
(and as usual the bilinear singlet mass parameter, $M_S^2$, see,
\eg, Eq.~(2.5) in \citere{Ellwanger:2009dp}).  The absolute value
of $\de A_\la$ is determined by the condition for on-shell charged Higgs
bosons.  The renormalization of $|A_\ka|$ is fixed by the triple \CP-even
singlet vertex, see \citere{NMSSMCT}.  The other cNMSSM-specific
input parameters are taken to be \DRbar\ parameters, see \eg \citere{NMSSMCT}.

We denote the complex input parameters as:
\begin{align}  
z = |z|\, e^{i\,\varphi_z} \qquad  \text{with} \qquad z = \la, \ka, \mueff, A_{\ka}, A_t\,.
\end{align}
The \DRbar\ renormalized parameters $\la$, $\ka$, $A_{\ka}$, $\TB$,
and $\mueff$ are then converted accordingly into \OS\ parameters:
\begin{align}
\la^{\OS} &= \la + \beta_\la \, \ln(Q^2/\mt^2)\,, \\
\ka^{\OS} &= \ka + \beta_\ka \, \ln(Q^2/\mt^2)\,, \\
A_{\ka}^{\OS} &= A_{\ka} + \beta_{A_{\ka}} \, \ln(Q^2/\mt^2)\,, \\  
\TB^{\OS} &= \TB + \beta_{\TB} \, \ln(Q^2/\mt^2)\,, \\
\mueff^{\OS} &= \mueff + \beta_{\mueff} \, \ln(Q^2/\mt^2)\,,  
\end{align}
with
\begin{align}
\beta_\la &= -\frac{\la}{32\, \pi^2\, \SW^2 \, \CW^2}
             \LB \al \, \pi \, (12\, \CW^2 + 4\, \SW^2) -
             \SW^2 \, \CW^2 \, (4\, |\la|^2 + 2\, |\ka|^2) + Y_\la \RB\,, \\
\beta_\ka &= \frac{3\, \ka}{16\, \pi^2}\, (|\la|^2 + |\ka|^2)\,, \\
\beta_{A_{\ka}} &= \frac{3}{8\, \pi^2}\, |\la|^2 \, \LB {|A_{\ka}}|^2\,
                  |\ka|^2/|\la|^2 + |A_\la|\,
                  \cos(\phi_{A_\la} - \phi_{A_\ka}) \RB\,, \\                    
\beta_{\TB} &= \frac{\TB}{32\, \pi^2}\, Y_{\TB}\,, \\
\beta_{\mueff} &= \mueff\, (\beta_\la/\la - \beta_\ka/(3\, \ka))\,, \\
Y_{\la} &= \sum_{g=1}^{3} \LB 3\, Y_{d_g}^2 + 3\, Y_{u_g}^2 + Y_{e_g}^2 \RB\,, \qquad
Y_{\TB} = \sum_{g=1}^{3} \LB 3\, Y_{d_g}^2 - 3\, Y_{u_g}^2 + Y_{e_g}^2 \RB\,, \\
Y_{d_g} &= \frac{e}{\sqrt{2} \MW \SW \CB}\, m_{d_g}\,, \quad
Y_{u_g} = \frac{e}{\sqrt{2} \MW \SW \SB}\, m_{u_g}\,, \quad
Y_{e_g} = \frac{e}{\sqrt{2} \MW \SW \CB}\, m_{e_g}\,,\\  
Q &= \mh1 + \MZ\,.
\end{align}
However, internally as default, \DRbar\ parameters and their
corresponding counter terms are used. It should furthermore be noted
that the phase of $A_{\ka}$ is left as a free parameter.

\bigskip
\subsubsection*{The MSSM limit}

We define the MSSM-limit of the NMSSM as
$\la \to 0$, $\ka \to 0$ with $\la/\ka$ fixed \ie $\la/\ka = 1$), and 
$\phi_H \to 0$, where the latter denotes a possible phase of the second
Higgs doublet.

%%%%%%%%%%%%%%%%%%%%%%%%%%%%%%%%%%%%%%%%%%%%%%%%%%%%%%%%%%%%%%%%%%%%%%%%%%%%%%%
%%%%%%%%%%%%%%%%%%%%%%%%%%%%%%%%%%%%%%%%%%%%%%%%%%%%%%%%%%%%%%%%%%%%%%%%%%%%%%%

\section{Calculation of diagrams}
\label{sec:calc}

In this section we give some details regarding the calculation of the
tree-level and higher-order corrections to the production of the lightest
Higgs boson $h_1 \equiv \hotf$ in $e^+e^-$ collisions, see \refeq{eq:eeh1Z}.
The diagrams and corresponding amplitudes have been obtained with 
\FA\ (version 3.11)~\cite{feynarts}, using the (N)MSSM model file%
\footnote{
  It should be noted that the NMSSM model file (including the corresponding
  driver files for \FC) which we have developed, are not yet officially part
  of \FA\ and \FC.
}
(including the (N)MSSM counterterms) of \citeres{MSSMCT, NMSSMCT}. 
The further evaluation has been performed with \FC\ (version 9.10) and 
\LT\ (version 2.16)~\cite{formcalc}.  The Higgs masses have been evaluated
using the Fortran program \texttt{NMSSMCALC} (version 5.0)~\cite{nmssmcalc}
with $\order{\al_t \al_s + \al_t^2}$ two-loop corrections and OS
renormalization in the (s)top-sector. 
Furthermore we use $\MHp$ as input parameter and allow for
general $\CP$-violation.
The numerical input values are given in \refse{sec:paraset}.

%%%%%%%%%%%%%%%%%%%%%%%%%%%%%%%%%%%%%%%%%%%%%%%%%%%%%%%%%%%%%%%%%%%%%%%%%%%%%%%

\subsection{Contributing diagrams}
\label{sec:diagrams}

% Fig1
%%%%%%%%%%%%%%%%%%%%%%%%% F I G U R E %%%%%%%%%%%%%%%%%%%%%%%%%%%%%%%%%%%%%%%%%
\begin{figure}
\begin{center}
\framebox[14.5cm]{\includegraphics[width=0.11\textwidth]{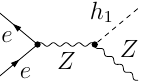}}\\
\framebox[14.5cm]{\includegraphics[width=0.73\textwidth]{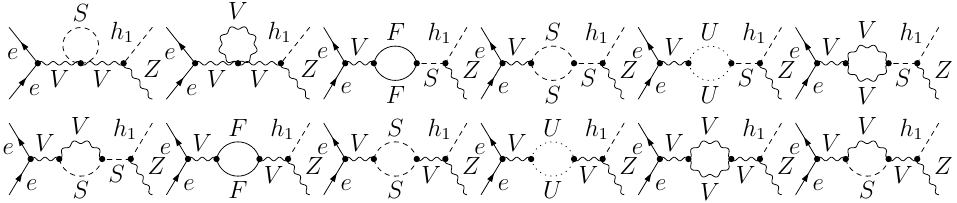}}\\
\framebox[14.5cm]{\includegraphics[width=0.73\textwidth]{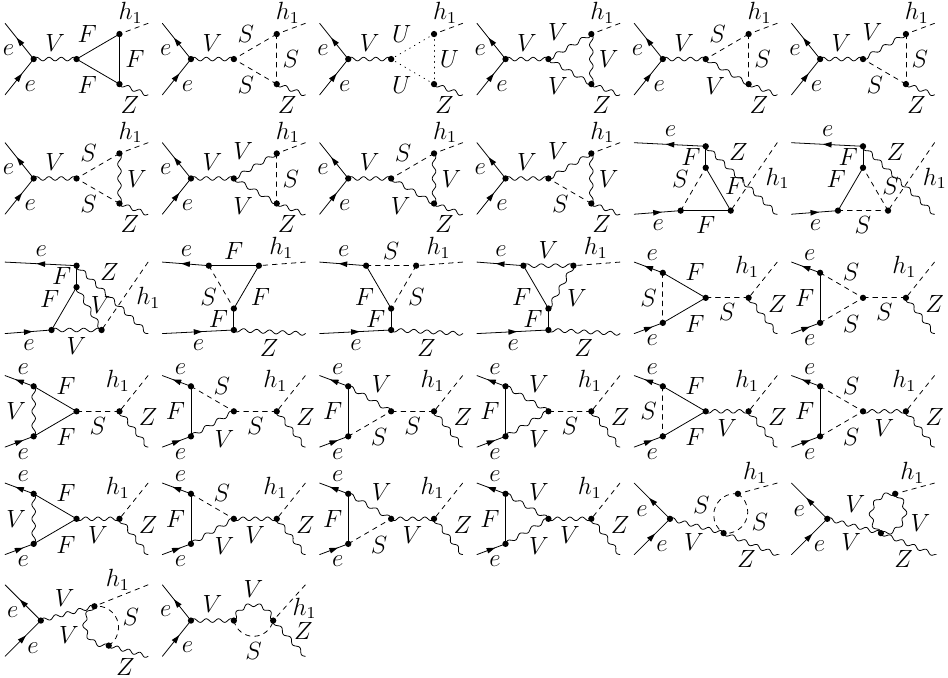}}\\
\framebox[14.5cm]{\includegraphics[width=0.73\textwidth]{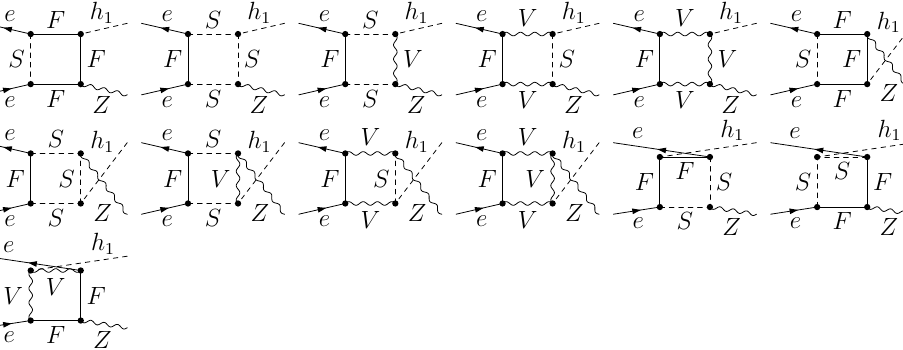}}\\
\framebox[14.5cm]{\includegraphics[width=0.71\textwidth]{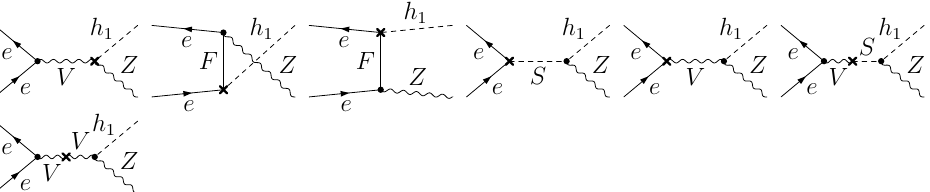}}
\caption{
  Generic tree, self-energy, vertex, box, and counterterm diagrams 
  for the process $e^+e^- \to h_1 Z$. 
  $F$ can be a SM fermion, chargino or neutralino; 
  $S$ can be a sfermion or a Higgs/Goldstone boson; 
  $V$ can be a $\ga$, $Z$ or $W^\pm$. 
  It should be noted that electron--Higgs couplings are neglected.  
}
\label{fig:hZdiagrams}
\end{center}
\end{figure}
%%%%%%%%%%%%%%%%%%%%%%%%% F I G U R E %%%%%%%%%%%%%%%%%%%%%%%%%%%%%%%%%%%%%%%%%

Sample diagrams for the process $e^+e^- \to h_1 Z$ are shown in  
\reffi{fig:hZdiagrams}.
Not shown are the diagrams for real (hard and soft) photon radiation. 
They are obtained from the corresponding tree-level diagrams by 
attaching a photon to the electrons/positrons.
The internal particles in the generically depicted diagrams in 
\reffi{fig:hZdiagrams} are labeled as follows: 
$F$ can be a SM fermion $f$, chargino $\cha{c}$ or neutralino 
$\neu{n}$; $S$ can be a sfermion $\Sf_s$ or a Higgs (Goldstone) boson 
$h_i, H^\pm$ ($G, G^\pm$); $U$ denotes the ghosts $u_V$;
$V$ can be a photon $\ga$ or a massive SM gauge boson, $Z$ or~$W^\pm$. 
As the impact of the electron--Higgs coupling on the computed
cross-sections is well below the theoretical uncertainty, we 
have neglected all these couplings via the \FA\ 
command \cite{feynarts}
\begin{align*}
\Code{Restrictions -> {NoElectronHCoupling}}
\end{align*}
and terms proportional to the electron mass \Code{ME}
(and the squared electron mass \Code{ME2}) via the \FC\ command 
\cite{formcalc}
\begin{align*}
\Code{Neglect[ME] = Neglect[ME2] = 0}~,
\end{align*}
which allows \FC\ to replace \Code{ME} by zero whenever this is safe, 
\ie except when it appears in negative powers or in loop integrals.
We have verified numerically that these contributions are indeed totally 
negligible.  For internally appearing Higgs bosons no higher-order
corrections to their masses or couplings are taken into account; 
these corrections would correspond to effects beyond one-loop order
for the process $e^+e^- \to h_1 Z$.%
\footnote{
  We found that using loop corrected Higgs boson masses 
  in the loops leads to a UV divergent result.
}
For external Higgs bosons, on-shell (OS) masses (including two-loop
corrections at $\order{\al_t \al_s + \al_t^2}$) are used, obtained
with \texttt{NMSSMCALC} (version 5.0)~\cite{nmssmcalc}.

Also not shown are the wave function correction diagrams with a
$Z$/Goldstone/Higgs--Higgs boson self-energy contribution on the external
$h_1$-boson leg.
They appear diagrammatically with a $Z/G/h_i$--$h_1$ transition
($i=1,\ldots,5$).
The $Z/G$--$h_1$ self-energy diagrams belonging to the process
$e^+e^- \to Z h_1$, yield a vanishing contribution for external on-shell
gauge bosons due to $\epsilon \cdot p = 0$ for $p^2 = \MZ^2$, where $p$ denotes
the external momentum and $\epsilon$ the polarization vector of the gauge boson.
The $h_1$--$h_1$-boson transition cannot be taken into account diagrammatically
in \FA/\FC\ because it leads to a division by zero.
Furthermore, the $h_1$--$h_i$ ($i \neq 1$) transition can only
be calculated at the \texttt{Particles} level in \FA/\FC, leading to
548 diagrams. Therefore we have implemented these self-energies directly via
\begin{align}
\hat{\Sigma}_{h_1h_i}(p^2) &= -\frac{2}{m_{h_{1}}^2- m_{h_{i}}^2}\,
                             \LV \Sigma_{h_1h_i}(p^2) + \tfrac{1}{2}\,
                             (m_{h_{1}}^2 - \bar{m}_{h_{1}}^2 +
                             m_{h_{1}}^2- \bar{m}_{h_{i}}^2)\, \de Z_{h_1h_i}
                             - \de M_{h_1h_i} \RV\,, 
\end{align}
with $ i \neq 1$ and $\bar{m}_{h_{i}}$ denotes the tree level Higgs masses,
while $m_{h_{i}}$ denotes the two-loop corrected Higgs masses at
$\order{\al_t \al_s + \al_t^2}$, obtained with \texttt{NMSSMCALC}
(version 5.0)~\cite{nmssmcalc}.
The diagonal contribution, \ie $i = 1$, is implemented directly by
the corresponding limit as (see, \eg, Eq.~(49) in \citere{Domingo:2021kud}),
\begin{align}
  \hat{\Sigma}^{\prime}_{h_1h_1}(p^2) &=
  -\frac{\partial}{\partial p^2}\Sigma_{h_1h_1}(p^2)
                                       -\de Z_{h_1h_1}\,.
\end{align}

Furthermore, in general, in \reffi{fig:hZdiagrams}
we have omitted diagrams with self-energy type corrections of external 
(on-shell) particles.  While the contributions from the real parts of the 
loop functions are taken into account via the renormalization constants 
defined by OS renormalization conditions, the contributions coming from 
the imaginary part of the loop functions can result in an additional (real) 
correction if multiplied by complex parameters.  In the analytical and 
numerical evaluation, these diagrams have been taken into account via the 
prescription given in \citere{MSSMCT}.

Within our one-loop calculation we neglect finite width effects that 
can help to cure threshold singularities.  Consequently, in the close 
vicinity of those thresholds our calculation does not give a reliable
result.  Switching to a complex mass scheme \cite{complexmassscheme} 
would be another possibility to cure this problem, but its application 
is beyond the scope of our paper.

For completeness we show here the tree-level cross section formula:
\begin{align}
\label{eq:eehZTree}
\sigma_{\text{tree}}(e^+e^- \to h_1 Z)  &= 
       \frac{\pi\, \alpha^2}{96\, s} 
       \LP \frac{8\, \SW^4 - 4\, \SW^2 + 1}{\SW^4\, \CW^4} \RP\,
       \frac{\la(1,\mh{1}^2/s,\MZ^2/s) + 12\,\MZ^2/s }{(1-\MZ^2/s)^2}\, 
\times \notag \\
&\mathrel{\phantom{=}}
       \la^{1/2}(1,\mh{1}^2/s,\MZ^2/s)\,
       \left| \CB\, Z^{\text{mix}}_{11} + \SB\, Z^{\text{mix}}_{12} \right|^2\,,
\end{align}
where $\la(x,y,z) = (x - y - z)^2 - 4yz$ denotes the usual two-body phase
space function and $Z^{\text{mix}}_{ij}$ denote the coefficients of the
transition matrix $\bold{Z}^{\text{mix}}$ to the physical Higgs fields,
which is further outlined in \citere{NMSSMCT}.

%%%%%%%%%%%%%%%%%%%%%%%%%%%%%%%%%%%%%%%%%%%%%%%%%%%%%%%%%%%%%%%%%%%%%%%%%%%%%%%

\subsection{Ultraviolet divergences}

As regularization scheme for the UV divergences we have used constrained 
differential renormalization~\cite{cdr}, which has been shown to be 
equivalent to dimensional reduction~\cite{dred} at the \onel\ 
level~\cite{formcalc}. 
Thus the employed regularization scheme preserves SUSY~\cite{dredDS,dredDS2}
and guarantees that the SUSY relations are kept intact, \eg that the gauge 
couplings of the SM vertices and the Yukawa couplings of the corresponding 
SUSY vertices also coincide to \onel\ order in the SUSY limit
(\ie for SUSY masses being equal to their corresponding SM masses).
Therefore no additional shifts, which might occur when using a different 
regularization scheme, arise. All UV divergences cancel in the final result.

%%%%%%%%%%%%%%%%%%%%%%%%%%%%%%%%%%%%%%%%%%%%%%%%%%%%%%%%%%%%%%%%%%%%%%%%%%%%%%%

\subsection{Infrared divergences}

Soft photon emission implies numerical problems in the phase space 
integration of radiative processes.  The phase space integral diverges 
in the soft energy region where the photon momentum becomes very small,
leading to infrared (IR) singularities.  Therefore the IR divergences from 
diagrams with an internal photon have to cancel with the ones from the 
corresponding real soft radiation.  We have included the soft photon contribution 
via the code already implemented in \FC\ following the description given 
in \citere{denner}.  The IR divergences arising from the diagrams involving 
a photon are regularized by introducing a photon mass parameter,
$\la_\ga$. All IR divergences,
\ie all divergences in the limit $\la_\ga \to 0$, cancel 
once virtual and real diagrams for one process are added. 
We have (numerically) checked that our results do not depend on $\la_\ga$.

\begin{figure}[t!]
\centering
\includegraphics[width=0.49\textwidth,height=7.5cm]{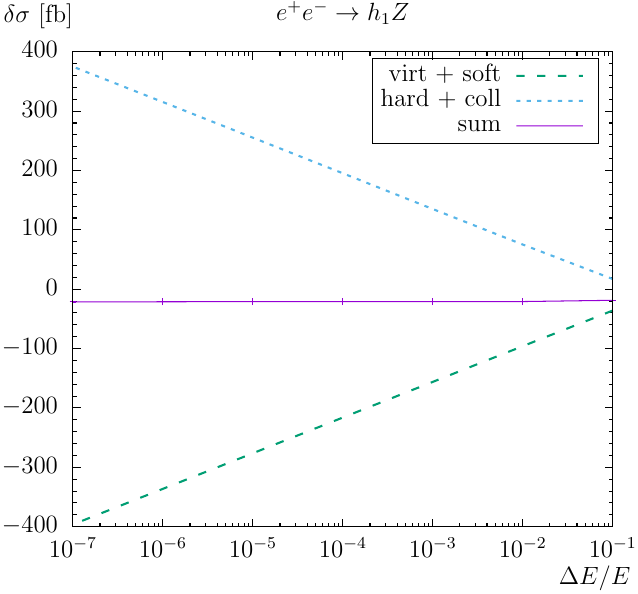}
\vspace{1em}
\begin{minipage}[b]{0.4\textwidth}
\centering
\begin{tabular}[b]{lr}
\toprule
$\Delta E/E$ & $\delta\sigma$/fbarn \\
\midrule
$10^{-1}$ & $-18.70 \pm 0.01$ \\
$10^{-2}$ & $-20.77 \pm 0.03$ \\
$10^{-3}$ & $-20.94 \pm 0.05$ \\
$10^{-4}$ & $-20.98 \pm 0.08$ \\
$10^{-5}$ & $-21.00 \pm 0.10$ \\
$10^{-6}$ & $-21.12 \pm 0.14$ \\
$10^{-7}$ & $-21.53 \pm 0.24$ \\
\bottomrule
\end{tabular}
\vspace{2em}
\end{minipage}
\includegraphics[width=0.49\textwidth,height=7.5cm]{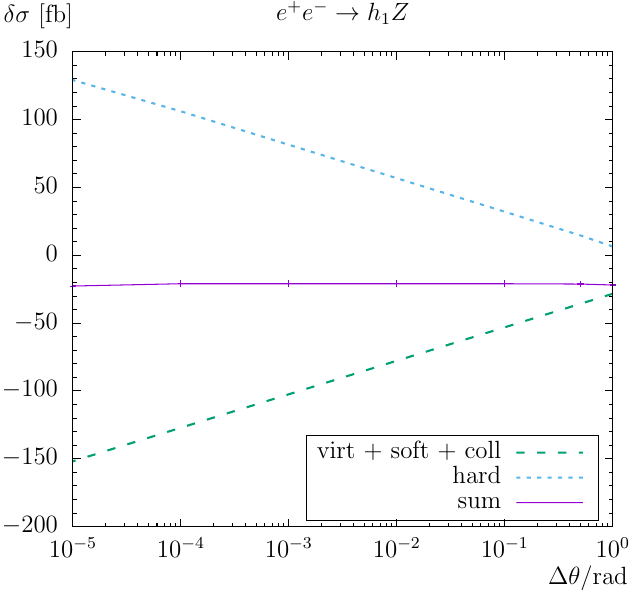}
\begin{minipage}[b]{0.4\textwidth}
\centering
\begin{tabular}[b]{lr}
\toprule
$\Delta \theta$/rad & $\delta\sigma$/fbarn \\
\midrule
$10^{ 0}$ & $-21.89 \pm 0.03$ \\
$10^{-1}$ & $-20.94 \pm 0.04$ \\
$10^{-2}$ & $-20.94 \pm 0.05$ \\
$10^{-3}$ & $-20.94 \pm 0.08$ \\
$10^{-4}$ & $-21.00 \pm 0.10$ \\
$10^{-5}$ & $-22.66 \pm 0.12$ \\
\bottomrule
\end{tabular}
\vspace{2em}
\end{minipage}
\caption{\label{fig:coll} 
  Phase space slicing method.  The different contributions to the loop 
  corrections $\delta\sigma(e^+e^- \to h_1 Z)$ at $\sqrt{s} = 250 \gev$
  with fixed $\Delta \theta/\text{rad} = 10^{-2}$ (upper plot) and
  fixed $\Delta E/E = 10^{-3}$ (lower plot).
}
\end{figure}

We have also numerically checked that our results do not depend on 
$\Delta E = \delta_s E = \delta_s \sqrt{s}/2$ defining the energy 
cut that separates the soft from the hard radiation.  As one can see
from the example in the upper plot of \reffi{fig:coll} this holds for 
several orders of magnitude.  Our numerical results below have been 
obtained for fixed $\delta_s = 10^{-3}$.

% Fig2
%%%%%%%%%%%%%%%%%%%%%%%%%% F I G U R E %%%%%%%%%%%%%%%%%%%%%%%%%%%%%%%%%%%%%%%
%%%%%%%%%%%%%%%%%%%%%%%%%% F I G U R E %%%%%%%%%%%%%%%%%%%%%%%%%%%%%%%%%%%%%%%

\subsection{Collinear divergences}

Numerical problems in the phase space integration of the radiative 
process arise also through collinear photon emission. Mass singularities 
emerge as a consequence of the collinear photon emission off massless
particles.  But already very light particles (such as electrons)
can produce numerical instabilities.

There are several methods for the treatment of collinear singularities. 
In the following, we give a very brief description of the so-called
\textit{phase space slicing} (PSS) \textit{method}~\cite{slicing}, 
which we adopted in our calculation.
The treatment of collinear divergences is not (yet) implemented in \FC, 
and therefore we have developed and implemented the code necessary for 
the evaluation of collinear contributions.

In the PSS method, the phase space is divided into
regions where the integrand is finite (numerically stable) and 
regions where it is divergent (or numerically unstable).
In the stable regions the integration is performed numerically, whereas
in the unstable regions it is carried out (semi-) analytically using 
approximations for the collinear photon emission~\cite{slicing}.

The collinear part is constrained by the angular cut-off parameter 
$\Delta\theta$, imposed on the angle between the photon and the
(in our case initial state) electron/positron.

The differential cross section for the collinear photon
that is radiated off from the initial state $e^+e^-$ pair corresponds to
a convolution 
\begin{align}
\text{d}\sigma_{\text{coll}}(s) = \frac{\alpha}{\pi} \int_0^{1-\delta_s} \text{d}z\,
  \text{d}\sigma_{\text{tree}}(\sqrt{z s}) \LV \LB 2\, \ln \LP 
  \frac{\Delta \theta \sqrt{s}}{2\, m_e} \RP - 1 \RB P_{ee}(z) + 1 - z \RV\,,
\end{align}
with $P_{ee}(z) = (1 + z^2)/(1 - z)$
denoting the splitting function of a photon from the initial $e^+e^-$ pair.
The electron momentum is reduced (because of the radiated photon) by 
the fraction $z$ such that the center-of-mass frame of the hard process 
receives a boost.  The integration over all possible factors $z$ is 
constrained by the soft cut-off $\delta_s = \Delta E/E$, to prevent 
over-counting in the soft energy region.

We have checked (numerically) that our results do not depend on 
$\Delta\theta$ over several orders of magnitude; see the example in 
the lower plot of \reffi{fig:coll}.  Our numerical results below have 
been obtained for fixed $\Delta \theta/\text{rad} = 10^{-2}$.

The one-loop corrections of the differential cross section are 
decomposed into the virtual, soft, hard, and collinear parts as follows:
\begin{align}
\text{d}\sigma_{\text{loop}} = 
  \text{d}\sigma_{\text{virt}}(\la) + 
  \text{d}\sigma_{\text{soft}}(\la, \Delta E) + 
  \text{d}\sigma_{\text{hard}}(\Delta E, \Delta\theta) + 
  \text{d}\sigma_{\text{coll}}(\Delta E, \Delta\theta)\,.
\end{align}
The hard and collinear parts have been calculated via the Monte Carlo
integration algorithm \texttt{Vegas} and \texttt{Divonne} as implemented in
the \texttt{CUBA} library \cite{cuba} as part of \FC.

%%%%%%%%%%%%%%%%%%%%%%%%%%%%%%%%%%%%%%%%%%%%%%%%%%%%%%%%%%%%%%%%%%%%%%%%%%%%%%%
%%%%%%%%%%%%%%%%%%%%%%%%%%%%%%%%%%%%%%%%%%%%%%%%%%%%%%%%%%%%%%%%%%%%%%%%%%%%%%%

\subsection{Tests and Comparisons}
\label{sec:comparisons}

In this section we present some tests and comparisons with results from
other papers for light neutral Higgs boson decay and production in
$e^+e^-$ collisions.

\begin{itemize}
\item     
  We have successfully checked our NMSSM tree-level cross section
  $\sigma_{\text{tree}}(e^+e^- \to h_1 Z)$
  against the analytical formula (\ref{eq:eehZTree}).
\item
  We have checked numerically that the MSSM-limit of the NMSSM
  (\ie $\la \to 0$, $\ka \to 0$ with $\la/\ka$ fixed (here
  $\la/\ka = 1$), $\phi_H \to 0$) holds.
  In particular, we checked that in the MSSM limit our results
  of \citere{NeutralHiggsProd} were found.
\item
  Furthermore we made some direct comparisons for $h_1 \to \tau^+ \tau^-$
  with \texttt{Mathematica} programs of \citere{NMSSMCT}, with full
  agreement.
\item
  The comparisons for the (light) neutral Higgs boson production within the
  MSSM can be found in \citere{NeutralHiggsProd}.
\item
  In \citere{Ham:2007mt} one-loop corrections to the Higgs sector were
  evaluated in the effective-potential approach, taking into account only
  (s)quarks of the third generation. The corresponding mixing was introduced
  into the effective $ZZh_1$~coupling. Furthermore they used a different
  input scheme and showed only scatter plots with
  $\sigma_0(hZZ) = \text{max}[\sigma(m_{h_1}), \sigma(m_{h_2}),
  \sigma(m_{h_3}), \sigma(m_{h_4}), \sigma(m_{h_5})]$, denoting the
  largest cross section among the five neutral Higgs bosons.

  The level of agreement of such a comparison depends on the correct
  transformation of the input scheme from our renormalization scheme into
  the schemes used in \citere{Ham:2007mt}, as well as on the differences in
  the employed renormalization schemes as such.
  In view of the different schemes and the non-trivial conversions such
  transformations and/or change of our renormalization prescription is
  beyond the scope of our paper. Furthermore, the fact that the
  results in \citere{Ham:2007mt} are only presented in scatter plots,
  renders a comparison unfeasible.
\item 
  In \citere{Cao:2014rma} only tree-level results (in scatter plots) are
  shown for the NMSSM, rendering a comparison unnecessary.
\end{itemize}

To our knowledge, there are no other papers dealing with (light) Higgs
boson production in $e^+e^-$ collisions within the NMSSM at the
\textit{full} one-loop level.

%%%%%%%%%%%%%%%%%%%%%%%%%%%%%%%%%%%%%%%%%%%%%%%%%%%%%%%%%%%%%%%%%%%%%%%%%%%%%
%%%%%%%%%%%%%%%%%%%%%%%%%%%%%%%%%%%%%%%%%%%%%%%%%%%%%%%%%%%%%%%%%%%%%%%%%%%%%

\section{Numerical analysis}
\label{sec:numeval}

In this section we present our numerical analysis of light neutral Higgs
boson production at $e^+e^-$ colliders in the Higgs-strahlung
channel in the c(N)MSSM. 
In the various figures below we show the cross sections at the tree level 
(``tree'') and at the full one-loop level (``full'').  

%%%%%%%%%%%%%%%%%%%%%%%%%%%%%%%%%%%%%%%%%%%%%%%%%%%%%%%%%%%%%%%%%%%%%%%%%%%%%

\subsection{Parameter settings}
\label{sec:paraset}

The renormalization scale $\mu_R$ has been set to the center-of-mass energy, 
$\sqrt{s}$.  The SM parameters are chosen as follows; see also \cite{pdg}:
\begin{itemize}

\item Fermion masses (on-shell masses, if not indicated differently):
\begin{align}
m_e    &= 0.51099895\mev\,,  & m_{\nu_e}    &= 0\,, \notag \\
m_\mu  &= 105.6583755\mev\,, & m_{\nu_{\mu}} &= 0\,, \notag \\
m_\tau &= 1776.86 \mev\,,    & m_{\nu_{\tau}} &= 0\,, \notag \\
m_u &= 71.8689\mev\,,        & m_d         &= 71.8689\mev\,, \notag \\ 
m_c &= 1.27\gev\,,           & m_s         &= 93.40\mev\,, \notag \\
m_t &= 172.69\gev\,,         & m_b         &= 4.18\gev\,.
\end{align}
According to \citere{pdg}, $m_s$ is an estimate of a so-called 
"current quark mass" in the \MSbar\ scheme at the scale 
$\mu \approx 2\gev$.  $m_c \equiv m_c(m_c)$ and $m_b \equiv m_b(m_b)$ 
are the "running" masses in the \MSbar\ scheme.  $m_u$ and $m_d$ are 
effective parameters, calculated through the hadronic contributions to
\begin{align}
\Delta\alpha_{\text{had}}^{(5)}(M_Z) &= 
      \frac{\alpha}{\pi}\sum_{f = u,c,d,s,b}
      Q_f^2 \Bigl(\ln\frac{M_Z^2}{m_f^2} - \frac 53\Bigr) \approx 0.02768\,.
\end{align}

\item Gauge-boson masses\index{gaugebosonmasses}:
\begin{align}
\MZ = 91.1876\gev\,, \qquad \MW = 80.377\gev\,.
\end{align}

\item Coupling constants\index{couplingconstants}:
\begin{align}
\al(0)  &= 1/137.03599918\,,  &\al(\MZ)  &= 1/127.951\,,  &\als(\MZ) &= 0.1179\,. 
\end{align}
\end{itemize}
The Higgs masses have been evaluated using \texttt{NMSSMCALC}
(version 5.0)~\cite{nmssmcalc}. 
For the corresponding MSSM masses the MSSM limit of the NMSSM is evaluated.

%%%%%%%%%%%%%%%%%%%%% T A B L E %%%%%%%%%%%%%%%%%%%%%%%%%%%%%%%%%%%%%%%%%%%%%%
\begin{table}[t!]
  \caption{\label{tab:para}
  NMSSM default input parameters in \DRbar\ 
  for the numerical investigation and the
  corresponding Higgs masses;
  all parameters (except of $\TB$, $\la$, $\ka$, $\phi_H$) are in
  GeV. Calculated masses (values) are rounded to 1 MeV ($10^{-3}$). 
  Also shown are the converted input parameters (\OS).
  The values for the trilinear sfermion Higgs couplings,
  $A_f$ are chosen such that charge- and/or color-breaking minima are
  avoided \cite{ccb}, with $A_{f\neq t} = 0$.
  It should be noted that for the first and second generation of sfermions we
  chose instead $M_{\tilde{Q}, \tilde{U}\!, \tilde{D}} = 3000\gev$.
  The phases of the potential complex parameters are zero in our
    benchmark scenario, i.e.\
    $\varphi_\la = \varphi_\ka = \varphi_{\mueff} = \varphi_{A_\ka}
    = \varphi_{\At} = 0$.
  $M_{h,H,A}$ denote the Higgs-boson masses. 
    $ZZh_1$ is the $Z$-$Z$-Higgs coupling in the NMSSM;
    $ZZh$ is the corresponding MSSM-limit,
    given by $ZZh = \frac{e\, \MW\, \SBA}{\CW^2\, \SW}$.
  $\CB\, Z^{\text{mix}}_{11} + \SB\, Z^{\text{mix}}_{12}$ is the
  ``NMSSM equivalent'' of $\SBA$ in the MSSM.
}
\centering
\begin{tabular}{rrrrrrrrrr}
\toprule  
$\sqrt{s}$ & sgn$(A_{\ka})$ & $\phi_H$ & $\MHp$ &
$M_{\tilde Q_3, \tilde U_3, \tilde D_3}$ & $M_{\tilde L, \tilde E}$ &
$|A_t|$ & $M_1$ & $M_2$ & $M_3$ \\
\midrule
250 & -1 & 0 & 1000 & 1500 & 1000 & 2000 & 600 & 400 & 2500 \\
\bottomrule
\end{tabular}

\vspace{0.5em}

\begin{tabular}{lrrrrrrr}
\toprule
Scen.  & $|\la|$ & $|\ka|$ & $|A_{\ka}|$ & $\TB$ & $|\mueff|$ \\
\midrule
\DRbar & 0.250 & 0.480 & 300.000 & 7.000 & 300.000 \\
\OS    & 0.251 & 0.481 & 301.299 & 6.972 & 300.754 \\  
\bottomrule
\end{tabular}

\vspace{0.5em}

\begin{tabular}{lrrrrrrrr}
\toprule
  Scen.  & $\mh1$ & $\mh2$ & $\mh3$ & $\mh4$ & $\mh5$ & $ZZh_1$ &
         $\CB\, Z^{\text{mix}}_{11} + \SB\, Z^{\text{mix}}_{12}$ \\
\midrule
\DRbar & 125.084 & 728.972 & 992.400 & 1000.919 & 1085.863 & 66.327472 &
                                                                  0.99992867 \\
\OS    & 125.072 & 731.186 & 992.457 & 1000.985 & 1087.438 & 66.327470 &
                                                                  0.99992865 \\
\bottomrule
\end{tabular}

\vspace{0.5em}

\begin{tabular}{lrrrrrrrr}
\toprule
Scen.  & $\Mh$ & $\MH$ & $\MA$ & $ZZh$ & $\SBA$ \\
\midrule
\DRbar & 125.062 & 996.828 & 996.575 & 66.332033 & 0.99999743 \\
\OS    & 125.043 & 996.831 & 996.575 & 66.332031 & 0.99999742 \\
\bottomrule
\end{tabular}

\end{table}
%%%%%%%%%%%%%%%%%%%%% T A B L E %%%%%%%%%%%%%%%%%%%%%%%%%%%%%%%%%%%%%%%%%%%%%%

\begin{figure}[t!]
\begin{center}
\begin{tabular}{c}
\includegraphics[width=0.48\textwidth,height=6cm]{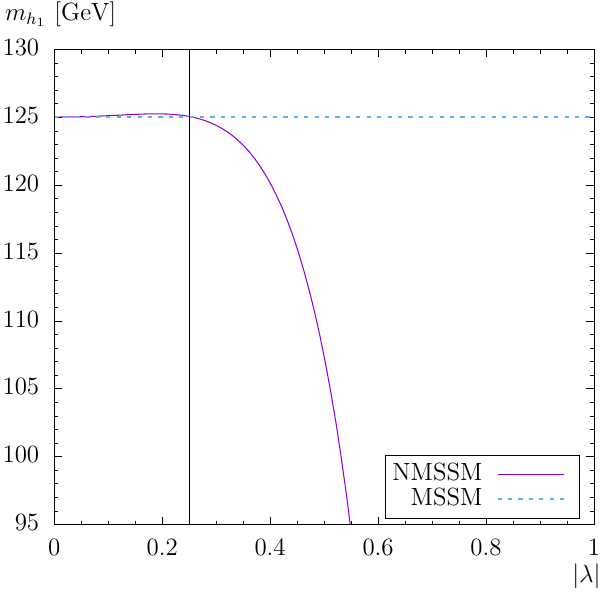}
\includegraphics[width=0.48\textwidth,height=6cm]{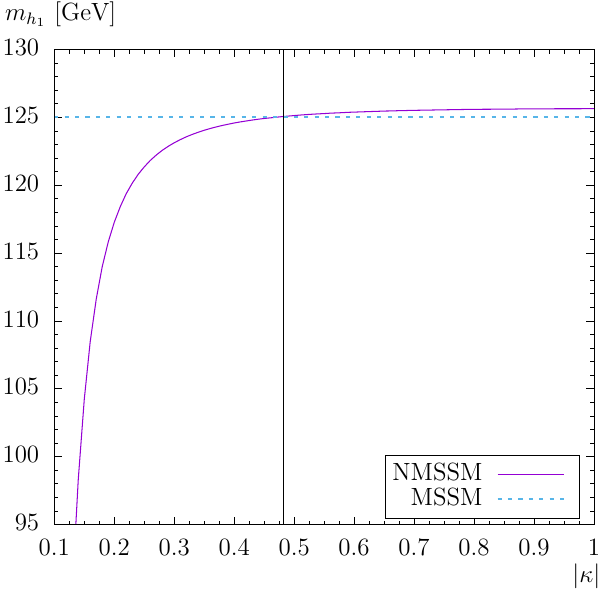}
\end{tabular}
\caption{\label{fig:Mh1}
  The light Higgs boson masses of the NMSSM and MSSM are shown with $|\la|$
  varied (left) and $|\ka|$ varied (right). The thin black lines cross
  with the purple solid and blue dotted lines in our default parameter
  point, see \refta{tab:para}.
}
\end{center}
\end{figure}

The SUSY parameters are chosen according to the scenario \OS\ shown in 
\refta{tab:para}.
This scenario constitutes a viable scenario for the process
$e^+e^- \to h_1 Z$ in the c(N)MSSM.
We do not strictly demand that the lightest Higgs boson has a mass around
$125 \pm 1.5\gev$, although for parts of the parameter space
shown in our plots below this is given. This is illustrated in
\reffi{fig:Mh1}, where we show $\mh1$ as a function of $|\la|$ (left) and
$|\ka|$ (right) in our scenario \OS. One can observe that $\mh1 \sim 125 \gev$
for $0 \le |\la| \lsim 0.3$ and $|\ka| \gsim 0.4$, where the actual values
of $|\la|$ and $|\ka|$ are indicated by vertical lines. Consequently, these
ranges can be seen as in agreement with the experimental
measurements. These ranges can vary if other parameters that have a
large impact on the prediction of $\mh1$ (such as $\At$) are varied.
Also shown is the value of $\Mh$ in the MSSM that has been obtained
in the MSSM limit of the NMSSM result (see the discussion above) and is
found to be $\Mh \approx 125.05 \gev$.

Furthermore, in our baseline scenario \OS\ one finds
$h_1 \sim h$, and since the $ZZh$ coupling in the MSSM
is $\propto \SBA \to 1$ in the
decoupling limit, relatively large cross sections are found.
This, however, can be different for the variation of the baseline
scenario. We will show the variation with respect to
$\sqrt{s}$, $|\la|$, $|\ka|$, $|\mueff|$,
$\MHp$, $M_{\tilde Q_3}$, $M_{\tilde U_3}$, $|A_{\ka}|$, $\At$ and $\TB$.
as well as the phases of the complex parameters,
$\varphi_{\la}$, $\varphi_{\ka}$,
$\varphi_{\mueff}$, and $\phiAt$, the phase of $\At$.

When performing an analysis involving complex parameters it should be 
noted that the results for physical observables are affected only by 
certain combinations of the phases of the parameters $\mu$, 
the trilinear couplings $A_f$ and the gaugino mass parameters 
$M_{1,2,3}$~\cite{MSSMcomplphasen,SUSYphases}.
For simplicity, we chose to vary the phases of $\la$, $\ka$,
$\mueff$ and $\At$ independently.

% Fig3
%%%%%%%%%%%%%%%%%%%%%%%%%% F I G U R E %%%%%%%%%%%%%%%%%%%%%%%%%%%%%%%%%%%%%%%
%%%%%%%%%%%%%%%%%%%%%%%%%% F I G U R E %%%%%%%%%%%%%%%%%%%%%%%%%%%%%%%%%%%%%%%

\medskip
The numerical results shown in the next subsections are of course 
dependent on the choice of the SUSY parameters.  Nevertheless, they 
give an idea of the relevance of the full one-loop corrections.

%%%%%%%%%%%%%%%%%%%%%%%%%%%%%%%%%%%%%%%%%%%%%%%%%%%%%%%%%%%%%%%%%%%%%%%%%%%%%%

\subsection{\texorpdfstring{
    The process \boldmath{$e^+e^- \to h_1 Z$} in the (N)MSSM}
  {The process e+e- -> h Z}}
\label{sec:eehZ}

The results shown in this and the following subsection consist of ``tree'',
which denotes the tree-level value, and of ``full'', which 
is the cross section including \textit{all} one-loop corrections 
as described in \refse{sec:calc}. 
All these processes are of particular interest for all proposed
future $e^+e^-$ colliders (as emphasized in \refse{sec:intro}).

%%%%%%%%%%%%%%%%%%%%%%%%%%%%%%%%%%%%%%%%%%%%%%%%%%%%%%%%%%%%%%%%%%%%%%%%%%%%%%%

\subsubsection{\texorpdfstring{Full one-loop results for varying
    real parameters} 
  {Full one-loop results for varying real parameters}}
\label{real}
          
\begin{figure}[t!]
\begin{center}
\begin{tabular}{c}
\includegraphics[width=0.48\textwidth,height=6cm]{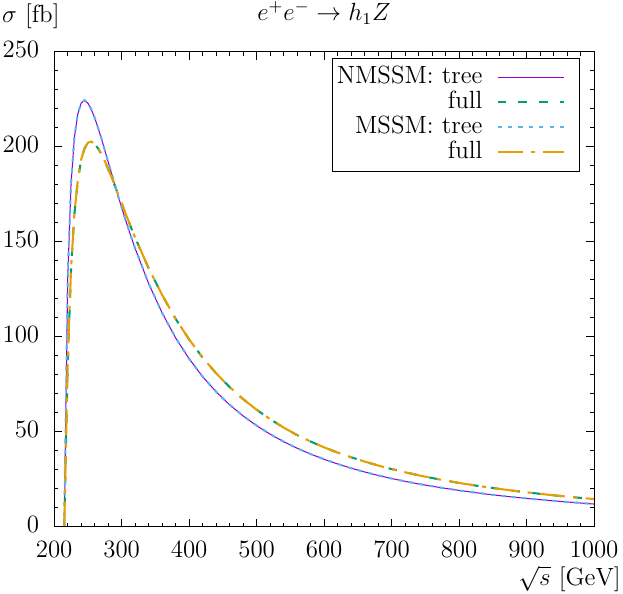}
\includegraphics[width=0.48\textwidth,height=6cm]{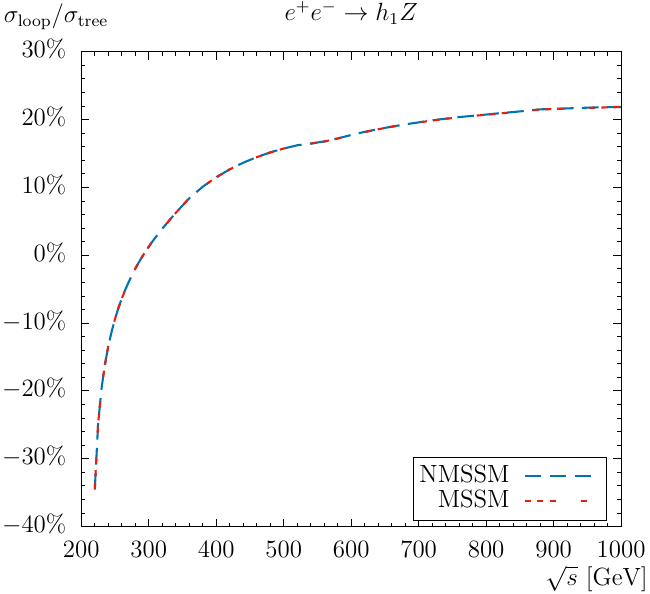}
\\[1em]
\includegraphics[width=0.48\textwidth,height=6cm]{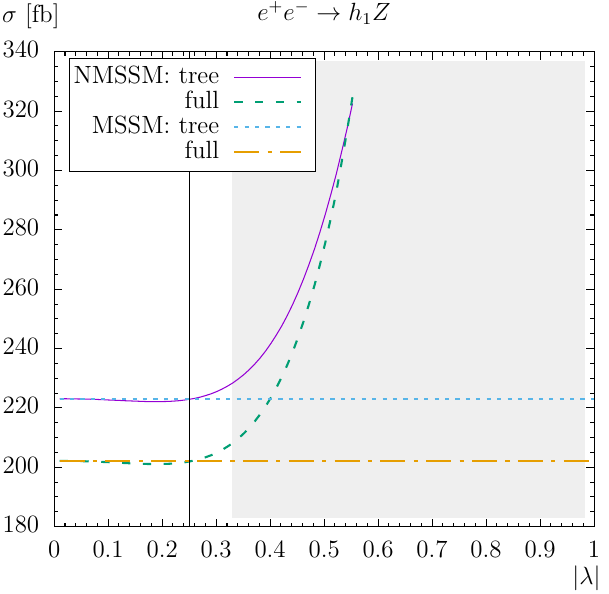}
\includegraphics[width=0.48\textwidth,height=6cm]{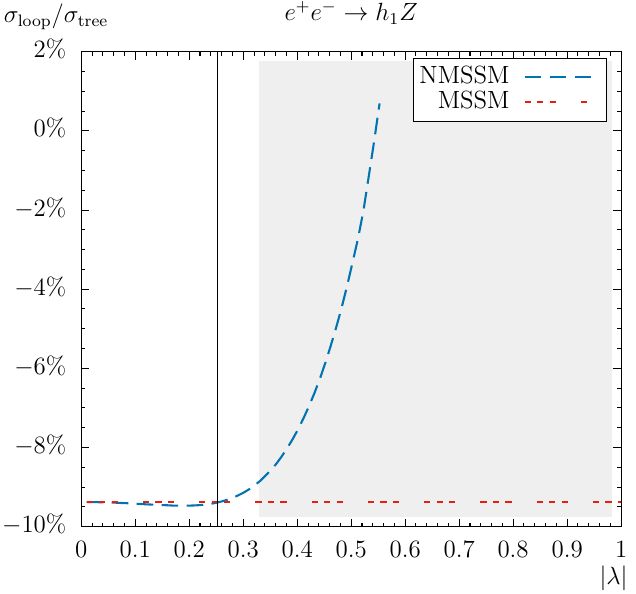}
\end{tabular}
\caption{\label{fig:eeh1Z_1}
  $\sigma(e^+e^- \to h_1 Z)$.
  Tree-level and full one-loop corrected cross sections in the (N)MSSM are
  shown with parameters chosen according to \refta{tab:para} (\OS). 
  The upper plots show the cross sections with $\sqrt{s}$ varied (left) and
  the corresponding $\sigma_{\mathrm{loop}}/\sigma_{\mathrm{tree}}$ in percent
  (right); the lower plots show $|\la|$ varied (left) and the
  corresponding $\sigma_{\mathrm{loop}}/\sigma_{\mathrm{tree}}$ in percent
  (right). The gray area is excluded by $\mh1 < 123.5 \gev$.
  The thin black solid line indicates $\mh1 \approx M_h \approx 125\gev$.
  It should be noted, that $\la \equiv 0$ in the MSSM limit, therefore here
  the MSSM cross section is constant.
}
\end{center}
\end{figure}

\begin{figure}[t!]
\begin{center}
\begin{tabular}{c}
\includegraphics[width=0.48\textwidth,height=6cm]{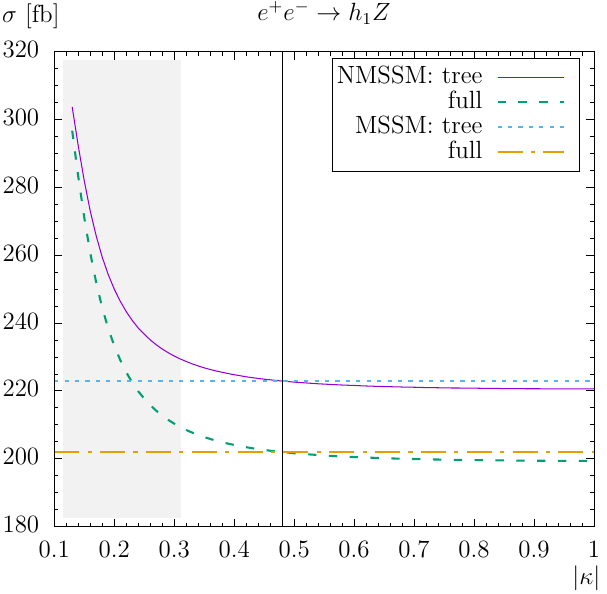}
\includegraphics[width=0.48\textwidth,height=6cm]{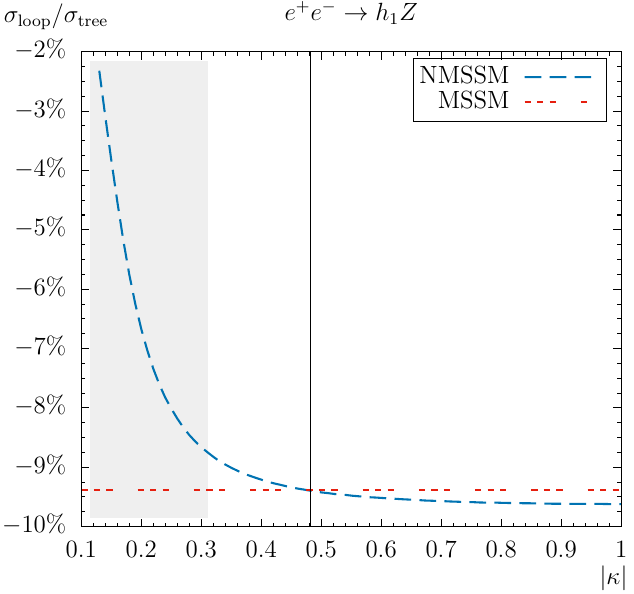}
\\[1em]
\includegraphics[width=0.48\textwidth,height=6cm]{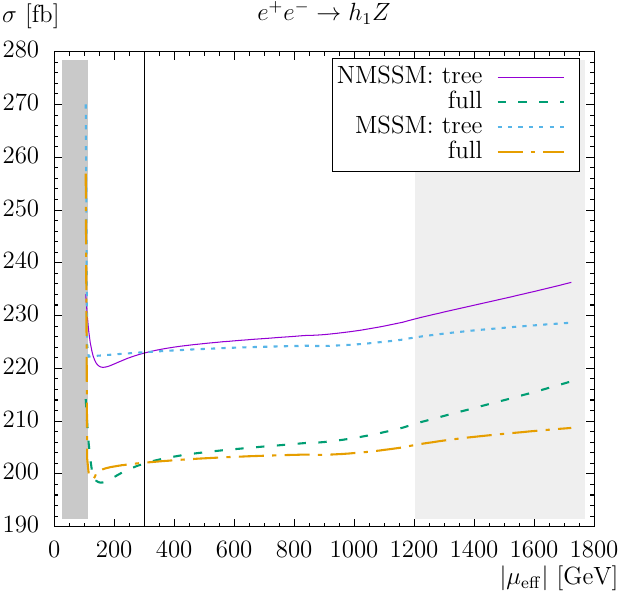}
\includegraphics[width=0.48\textwidth,height=6cm]{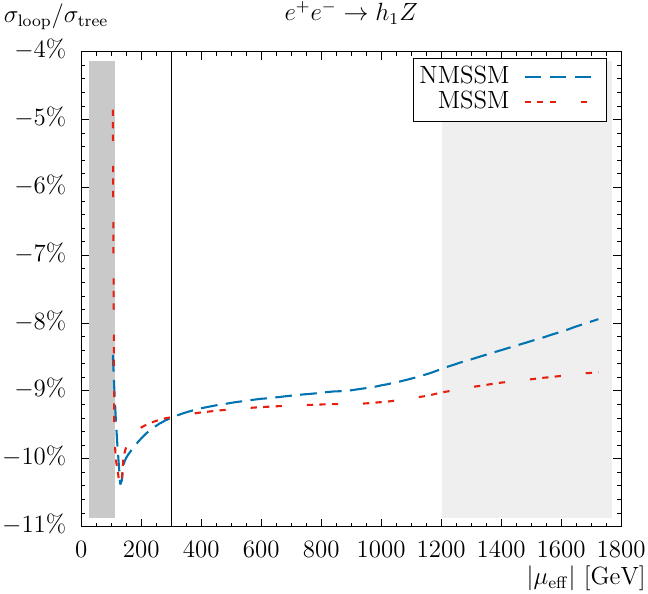}
\end{tabular}
\caption{\label{fig:eeh1Z_2}
  $\sigma(e^+e^- \to h_1 Z)$.
  Tree-level and full one-loop corrected cross sections in the (N)MSSM are
  shown with parameters chosen according to \refta{tab:para} (\OS).  
  The upper plots show the cross sections with $|\ka|$ varied (left) and
  the corresponding $\sigma_{\mathrm{loop}}/\sigma_{\mathrm{tree}}$ in percent
  (right). It should be noted, that $\ka \equiv 0$ in the MSSM limit,
  therefore, here the MSSM cross section is constant.
  The lower plots show $\mueff$ varied (left) and the corresponding
  $\sigma_{\mathrm{loop}}/\sigma_{\mathrm{tree}}$ in percent (right).
  The light-gray (dark-gray) areas are excluded by $\mh1 < 123.5 \gev$
  $(M_h < 123.5 \gev)$. The thin black solid line indicates
  $\mh1 \approx M_h \approx 125\gev$.
}
\end{center}
\end{figure}

\begin{figure}[t!]
\begin{center}
\begin{tabular}{c}
\includegraphics[width=0.45\textwidth,height=6cm]{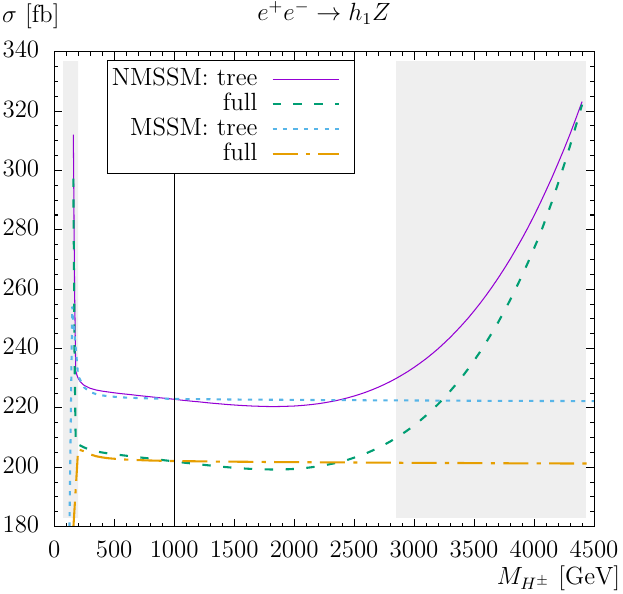}
\includegraphics[width=0.45\textwidth,height=6cm]{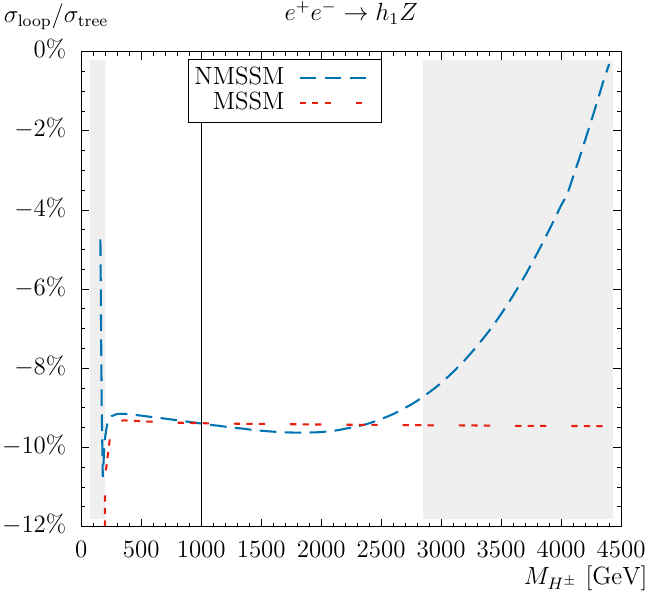}
\\[0.5em]   
\includegraphics[width=0.45\textwidth,height=6cm]{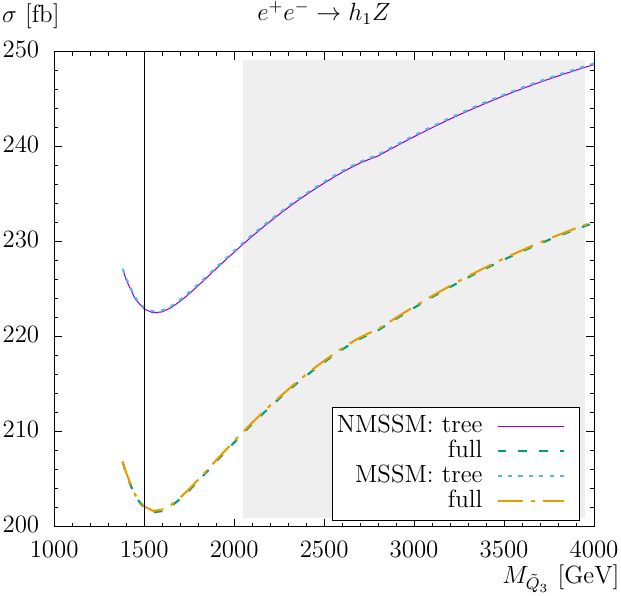}
\includegraphics[width=0.45\textwidth,height=6cm]{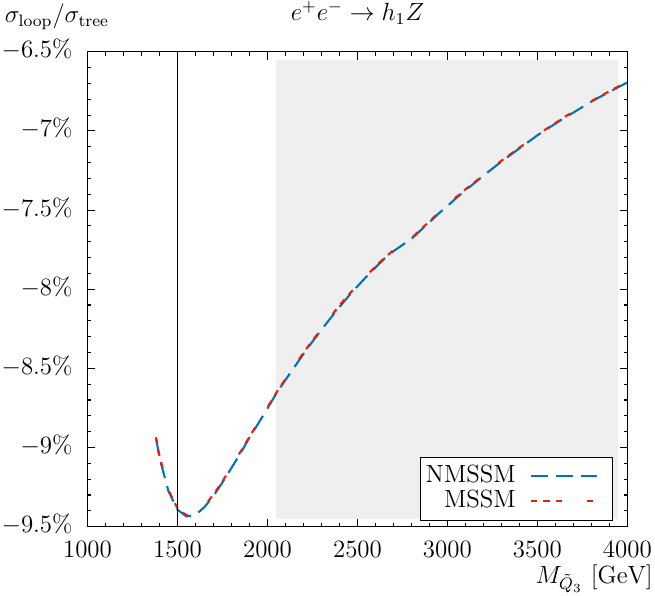}
\\[0.5em]
\includegraphics[width=0.45\textwidth,height=6cm]{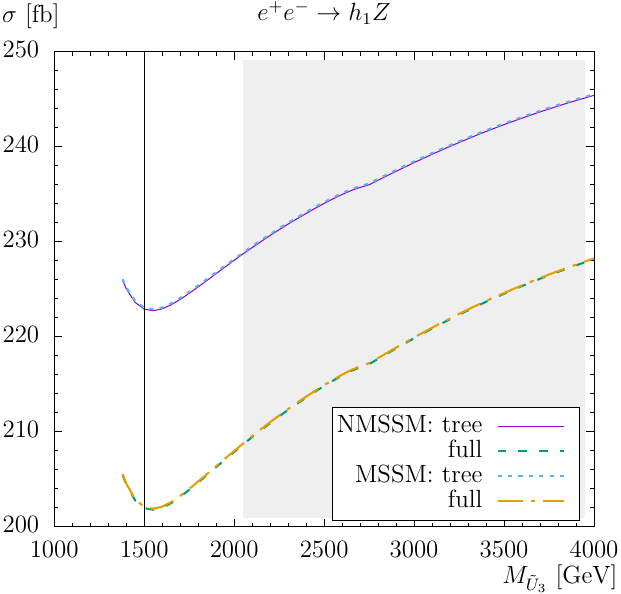}
\includegraphics[width=0.45\textwidth,height=6cm]{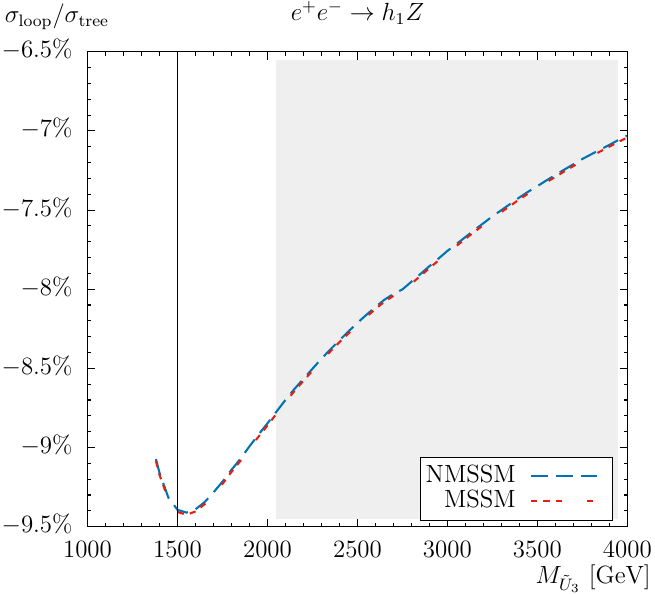}
\end{tabular}
\caption{\label{fig:eeh1Z_3}
  $\sigma(e^+e^- \to h_1 Z)$.
  Tree-level and full one-loop corrected cross sections in the (N)MSSM are
  shown with parameters chosen according to \refta{tab:para} (\OS).
  The upper plots show the cross sections with $\MHp$ varied (left) and the
  corresponding $\sigma_{\mathrm{loop}}/\sigma_{\mathrm{tree}}$ in percent
  (right).
  The middle plots show the cross sections with $M_{\tilde{Q}_3}$ varied (left)
  and the corresponding $\sigma_{\mathrm{loop}}/\sigma_{\mathrm{tree}}$ in
  percent (right).
  The lower plots show $M_{\tilde{U}_3}$ varied (left) and the corresponding
  $\sigma_{\mathrm{loop}}/\sigma_{\mathrm{tree}}$ in percent (right).
  The light-gray area is excluded by $\mh1 \approx M_h < 123.5 \gev$.
  The thin black solid line indicates $\mh1 \approx M_h \approx 125\gev$. 
}
\vspace{-2em}
\end{center}
\end{figure}

\begin{figure}[t!]
\begin{center}
\begin{tabular}{c}
\includegraphics[width=0.45\textwidth,height=6cm]{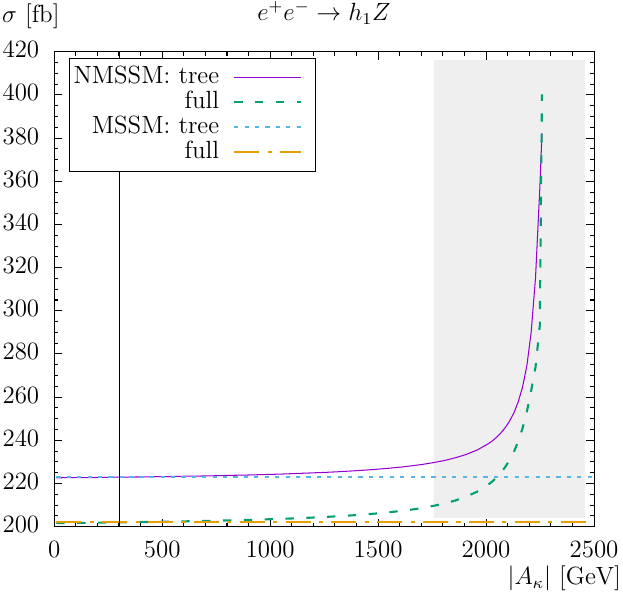}
\includegraphics[width=0.45\textwidth,height=6cm]{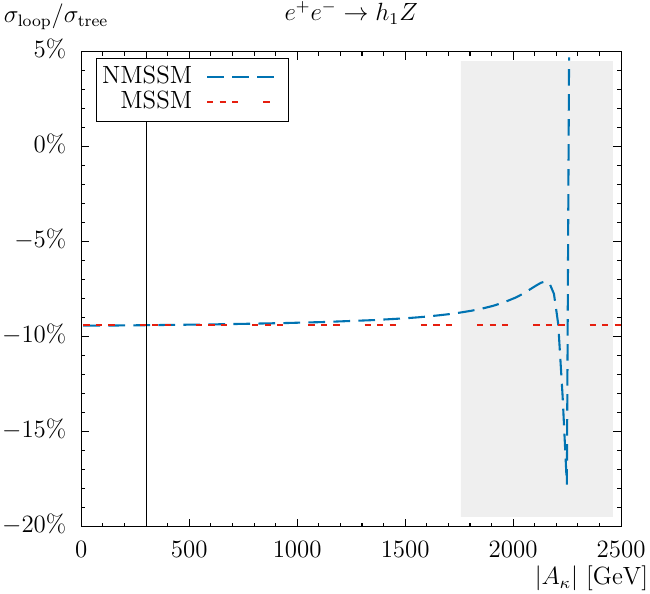}
\\[0.5em]
\includegraphics[width=0.45\textwidth,height=6cm]{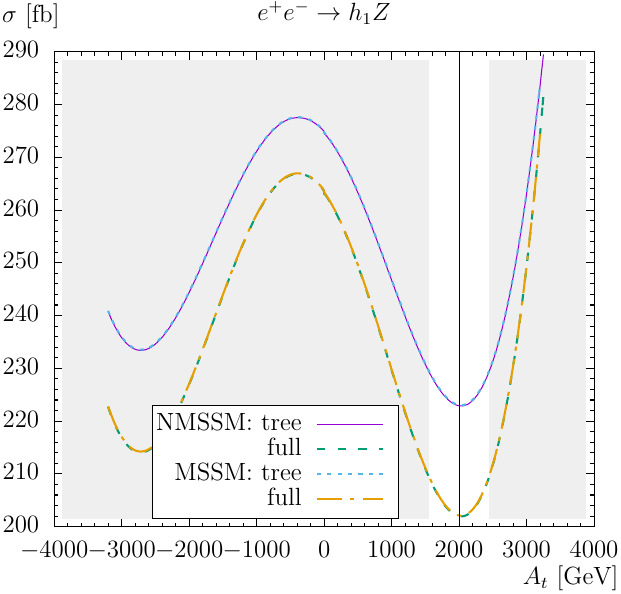}
\includegraphics[width=0.45\textwidth,height=6cm]{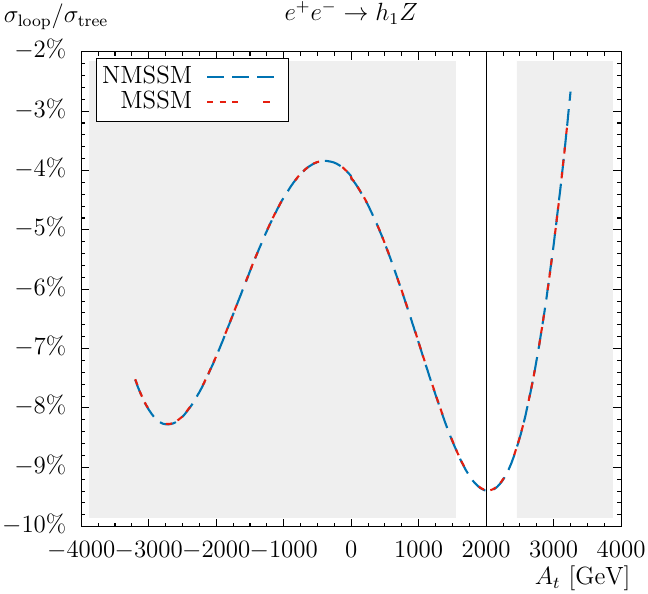}
\\[0.5em]
\includegraphics[width=0.45\textwidth,height=6cm]{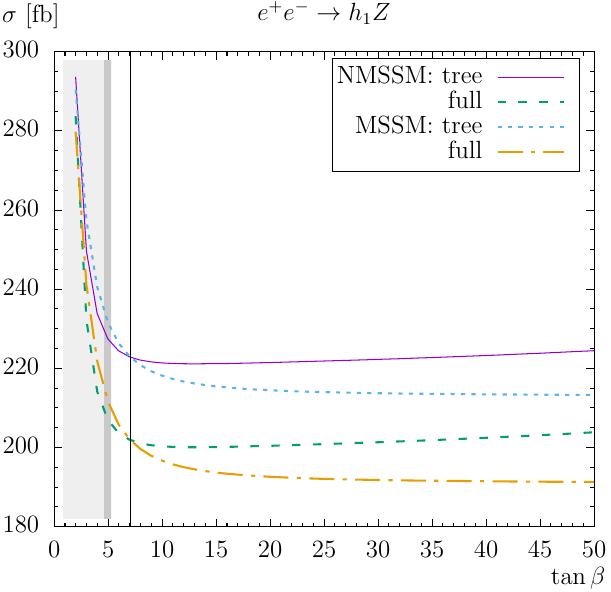}
\includegraphics[width=0.45\textwidth,height=6cm]{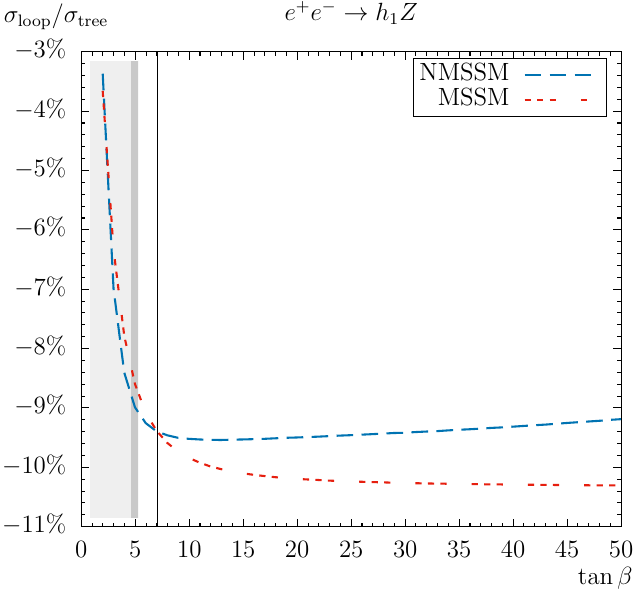}
\end{tabular}
\caption{\label{fig:eeh1Z_4}
  $\sigma(e^+e^- \to h_1 Z)$.
  Tree-level and full one-loop corrected cross sections in the (N)MSSM are
  shown with parameters chosen according to \refta{tab:para} (\OS).  
  The upper plots show the cross sections with $|A_{\ka}|$ varied (left)
  and the corresponding $\sigma_{\mathrm{loop}}/\sigma_{\mathrm{tree}}$ in
  percent (right).
  The middle plots show the cross sections with $A_t$ varied (left) and the
  corresponding $\sigma_{\mathrm{loop}}/\sigma_{\mathrm{tree}}$ in percent
  (right).
  The lower plots show $\TB$ varied (left) and the corresponding
  $\sigma_{loop}/\sigma_{tree}$ in percent (right).
  The light-gray (dark-gray) area is excluded by $\mh1 < 123.5 \gev$
  $(M_h < 123.5 \gev)$. The thin black solid line indicates
  $\mh1 \approx M_h \approx 125\gev$.
}
\vspace{-2em}
\end{center}
\end{figure}

We begin the numerical analysis with the cross sections of $e^+e^- \to h_1 Z$
evaluated as a function of the real parameters in the NMSSM.
In \reffis{fig:eeh1Z_1} -- \ref{fig:eeh1Z_4} we show the results for the 
process $e^+e^- \to h_1 Z$ in the (N)MSSM, as a function of $\sqrt{s}$,
$|\la|$, $|\ka|$, $|\mueff|$, $\MHp$, $M_{\tilde Q_3}$,
$M_{\tilde U_3}$, $|A_{\ka}|$, $\At$ and $\TB$, respectively.

\medskip

% Fig4
%%%%%%%%%%%%%%%%%%%%%%%%%% F I G U R E %%%%%%%%%%%%%%%%%%%%%%%%%%%%%%%%%%%%%%%
%%%%%%%%%%%%%%%%%%%%%%%%%% F I G U R E %%%%%%%%%%%%%%%%%%%%%%%%%%%%%%%%%%%%%%%

We start with the process $e^+e^- \to h_1 Z$ in the \OS\ scenario as a
function of $\sqrt{s}$ as shown in the upper part of \reffi{fig:eeh1Z_1}.
The total cross section starts at the production threshold
$\mh1+\MZ=216.260\gev$. At the tree-level one finds a maximum of about
$203\;\fb$ in the (N)MSSM at $\sqrt{s} \approx 255\gev$ with a decrease
for increasing $\sqrt{s}$. The results of the NMSSM are effectively
indistinguishable from the MSSM. 
The size of the corrections to the cross section (upper right plot in
\reffi{fig:eeh1Z_1}) can be especially large in close vicinity
to the production
threshold%
\footnote{
  It should be noted that a calculation very close to the production 
  threshold requires the inclusion of additional (nonrelativistic) 
  contributions, which is beyond the scope of this paper. 
  Consequently, very close to the production threshold our calculation 
  (at the tree and loop level) does not provide a very accurate 
  description of the cross section.
}
from which on the considered process is kinematically possible.
At the production threshold we found relative corrections
(\ie $\sigma_{\mathrm{loop}}/\sigma_{\mathrm{tree}}$) of 
$\approx -30\%$.  Away from the production threshold, loop corrections of 
$\approx -10\%$ at $\sqrt{s} = 250\gev$ are found, increasing to 
$\approx +22\%$ at $\sqrt{s} = 1000\gev$. 
Also at the one-loop level the deviation of
$\sigma_{\mathrm{loop}}/\sigma_{\mathrm{tree}}$
between the NMSSM and MSSM is extremely small, i.e.\ the loop
corrections do not differ in a relevant way due to the extended NMSSM
particle spectrum. It should be noted that this equality
  crucially depends on the fact that
the scenario results in effectively identical 
light Higgs boson masses in the two models, see \refta{tab:para}.

Following the running scenarios of the proposed future $e^+e^-$ colliders,
supported by the location of the maximum cross sections discussed above,
in the following plots we always assume $\sqrt{s} = 250 \gev$. 

\medskip

In the lower part of \reffi{fig:eeh1Z_1} the cross section is analyzed in
dependence of $|\la|$.%
\footnote{
Here and in the following we denote the potentially complex parameters $z$
as $|z|$, keeping in mind that in the benchmark scenario we have
defined $\varphi_z = 0$. We specify explicitely when we deviate
from $\varphi_z = 0$.
}
The left plot shows the cross section starting at
$|\la| \approx 0.01$ (the lower bound on $\mneu1$ is violated for
$|\la| < 0.01$). The part of the parameter space where $\mh1$ deviates
more than about $1.5 \gev$ from $\Mh$, see \reffi{fig:Mh1}, is shaded in
gray. As an example, starting from $|\la| \approx 0.3$
the NMSSM cross section
strongly increases up to $325\;\fb$ at $|\la| \approx 0.55$, where $\mh1
= 93.65\gev$. Consequently, this increase is a pure kinematic effect.
The relative corrections, as shown in the lower right plot, are nearly
constant in the allowed parameter space and reach $\approx -9.5\%$.
The MSSM limit is constant (because $\la \equiv 0$) at $202\;\fb$ and
the relative correction is found to be effectively identical to the NMSSM.

\medskip

% Fig5
%%%%%%%%%%%%%%%%%%%%%%%%%% F I G U R E %%%%%%%%%%%%%%%%%%%%%%%%%%%%%%%%%%%%%%%
%%%%%%%%%%%%%%%%%%%%%%%%%% F I G U R E %%%%%%%%%%%%%%%%%%%%%%%%%%%%%%%%%%%%%%%

In \reffi{fig:eeh1Z_2} we analyze the cross section in dependence of
$|\ka|$ (upper row) and $|\mueff|$ (lower row).
In the upper left plot we start at $|\ka| \approx 0.13$ because
$m^2_{h_1} < 0$ for lower values.
The excluded region for $\mh1 < 123.5\gev$ is in light-gray.
Again the MSSM limit is constant due to $\ka \equiv 0$ at $202\;\fb$, and the
relative correction is about $-9.5\%$, as in the case of the
$|\la|$ variation.  The maximum deviation of the relative corrections
between the NMSSM and MSSM amounts only to
$0.64\%$ at $|\ka| \approx 0.31$. However, these small differences
of up to $\sim 5\;\fb$ 
can be attributed to the deviation between $\mh1$ and $\Mh$ reaching
$1.5 \gev$. 

\medskip

In the lower left plot the NMSSM cross section is shown as a function of
$|\mueff|$, 
starting at $\approx 105\gev$ because $\mcha1 < 94\gev$ (the lower limit
for $|\mueff| < 103\gev$ in our scenario).
The light-gray (dark-gray) areas are excluded by $\mh1 < 123.5 \gev$
$(M_h < 123.5 \gev)$. The thin black solid line indicates
$\mh1 \approx M_h \approx 125\gev$.
The NMSSM cross section reaches a minimum of $198\;\fb$ at
$\mueff \approx 155\gev$ and
slightly increases with increasing $\mueff$. A smaller increase is
visible also for the MSSM results, and the small deviations between the
NMSSM and MSSM cross sections, reaching $\pm 5\;\fb$, can again be
attributed to changes in $\mh1$ of up to $\pm 1.5 \gev$. 
The relative corrections (lower right plot) in the (N)MSSM reach
about $-10.5\%$ in the respective minimum. 
The maximum deviation of the relative corrections between the NMSSM
and MSSM is (in the allowed region) $0.4\%$ at $|\mueff| \approx 1203\gev$,
effectively due to the change in $\mh1$. 

\medskip

% Fig6
%%%%%%%%%%%%%%%%%%%%%%%%%% F I G U R E %%%%%%%%%%%%%%%%%%%%%%%%%%%%%%%%%%%%%%%
%%%%%%%%%%%%%%%%%%%%%%%%%% F I G U R E %%%%%%%%%%%%%%%%%%%%%%%%%%%%%%%%%%%%%%%

In the upper left plot of \reffi{fig:eeh1Z_3} we show the (N)MSSM cross
sections as a function of $\MHp$. As before the regions with
$\mh1 < 123.5 \gev$ are light gray shaded, where the left shaded area
is also valid for the MSSM. Overall, the cross section is nearly
constant, with a small variation due to the change in $\mh1$ or $\Mh$.
The NMSSM loop corrections are mostly at the level of $-9.5\%$ in the
allowed area, with a similar result for the MSSM. 
The maximum deviation of the relative corrections between the NMSSM and
MSSM amounts to $\approx 0.2\%$ at $1800\gev$.

\medskip

The effects of $M_{\tilde Q_3}$ and $M_{\tilde U_3}$, shown in the middle and
lower plots of \reffi{fig:eeh1Z_3}, respectively, are quite similar, and we
restrict ourselves to describing the first of this soft breaking
parameters. Different values for $M_{\tilde U_3}$ are given in
brackets. Here, the 
gray shaded area indicates $\mh1 \approx \Mh < 123.5 \gev$, where the
two Higgs-boson masses exhibit the same dependence on the soft
SUSY-breaking parameters in the stop sector.
The cross sections start at $1380\gev$ because lower values violate the
lower bound of $\mstop1 < 1310\gev$.
The cross section is found to vary at the level of $\pm 5\;\fb$,
following the change in 
the light Higgs-boson masses and partially due to genuine scalar top
loops in the production cross section. The loop
corrections amount to $\approx -8.7\, (8.8)\%$ to $\approx -9.4\%$
for the (N)MSSM. The deviation of the relative corrections between the
NMSSM and MSSM is negligibly small. 

% Fig7
%%%%%%%%%%%%%%%%%%%%%%%%%% F I G U R E %%%%%%%%%%%%%%%%%%%%%%%%%%%%%%%%%%%%%%%
%%%%%%%%%%%%%%%%%%%%%%%%%% F I G U R E %%%%%%%%%%%%%%%%%%%%%%%%%%%%%%%%%%%%%%%

\medskip

In the upper row of \reffi{fig:eeh1Z_4} we show the dependence of the
Higgs-boson production cross sections on $|A_\ka|$.
Here it should be noted that $\MHp$ is one of our input parameters
and thus remains fixed when $|A_\ka|$ is varied.
As before, the light
gray region is excluded by $\mh1 < 123.5 \gev$.
The full MSSM cross section is constant in the allowed parameter space
at $222\;\fb$, with an extremely
small variation of the NMSSM cross section at larger values of $|A_\ka|$. 
The relative corrections (upper right plot) can reach $\approx -9.4\%$
in the allowed region, for both the MSSM and NMSSM. 
The deviation of the relative corrections between the NMSSM and MSSM
reach only $0.14\%$.

\medskip

The variation of the cross sections with $\At$ is shown in the middle
plots of \reffi{fig:eeh1Z_4}, with the gray area excluded by
$\mh1 \approx \Mh < 123.5 \gev$. 
First it should be noted that the structure around $-400\gev$ is caused
by the loop corrected Higgs masses from \texttt{NMSSMCALC} through a
conversion from $A_t^{\DRbar}$ to $A_t^{\OS}$ which does not converge
properly. Therefore a simplified conversion is used (\ie no iteration).
The cross section follows the mass dependence allowed by the
$\pm 1.5 \gev$ uncertainty around $125 \gev$, where the lowest cross
section is reached for the highest Higgs-boson mass of $\sim 125 \gev$,
indicated by the vertical black line. For lower or higher $A_t$ values
the Higgs-boson mass goes down, and the cross section rises accordingly,
i.e.\ the variation is a pure artefact of the change in $\mh1 \approx \Mh$.
The loop corrections in the allowed interval are about $-9.5\%$ in both
models, see the middle right plot.
The deviation of the relative corrections between the NMSSM and MSSM
is negligible.

\medskip

Finally, the dependence on $\TB$ is shown in the lower plots of
\reffi{fig:eeh1Z_4}. 
The NMSSM loop corrections can reach $\approx -9.5\%$ at $\TB = 13$, while
the MSSM loop corrections reach $\approx -10.3\%$ at $\TB = 50$.
The maximum deviation of the relative corrections between the NMSSM and MSSM
reach $\approx 1.1\%$ at $\TB = 50$, which is (as before) a pure kinematic
effect because of the different light Higgs masses.

\medskip

Overall, we observe that the prediction of the light Higgs-boson
production cross section in the NMSSM is nearly indistinguishable from
the MSSM cross section. Nearly all (small) observed effects are
kinematic, originating from the small variation allowed in $\mh1$ due to
the theoretical uncertainties, see \citere{Slavich:2020zjv}. However,
since the value of the Higgs-boson mass is determined experimentally to
a high accuracy, these kinematic effects must be considered numerical
artefacts.
The main genuine loop effects of the NMSSM and the MSSM were
observed at the percent level in the dependence on
$M_{\tilde Q_3}$, $M_{\tilde U_3}$ and $A_t$, where the differences
between the NMSSM and the MSSM are negligibly small.

%%%%%%%%%%%%%%%%%%%%%%%%%%%%%%%%%%%%%%%%%%%%%%%%%%%%%%%%%%%%%%%%%%%%%%%%%%%%%

\subsubsection{\texorpdfstring{Full one-loop results for varying phases}
{Full one-loop results for varying phases}}

We finish the numerical analysis with the cross sections of
$e^+e^- \to h_1 Z$ evaluated as a function of the phases of two
(potentially) complex parameters in the
(N)MSSM: $\varphi_{\la}$, $\varphi_{\ka}$, 
as well as $\varphi_{\mueff}$ and $\phiAt$ (all between $0^{\circ}$ and
$360^{\circ}$). As already emphasized, $\la$ and $\ka$ are zero in the
MSSM limit, therefore the MSSM cross section is constant.

\begin{figure}[ht!]
\begin{center}
\begin{tabular}{c}
\includegraphics[width=0.45\textwidth,height=6cm]{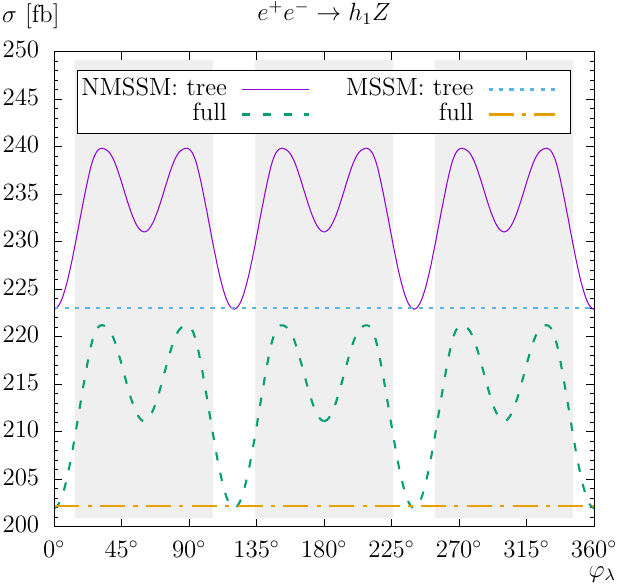}
\includegraphics[width=0.45\textwidth,height=6cm]{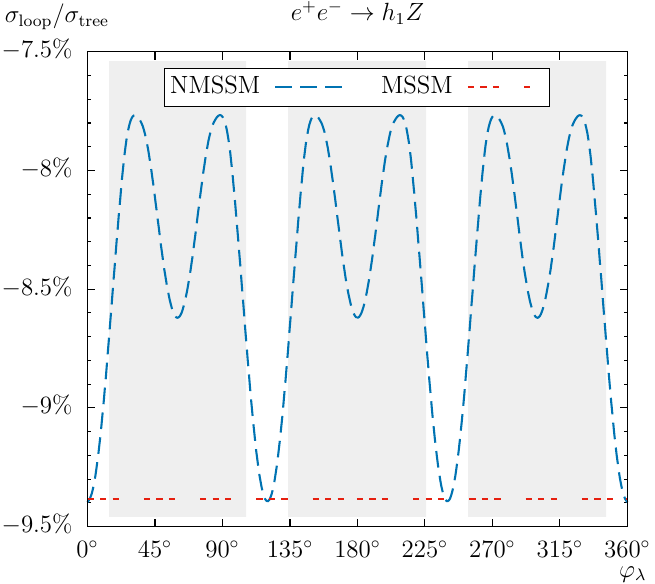}
\\[0.5em]
\includegraphics[width=0.45\textwidth,height=6cm]{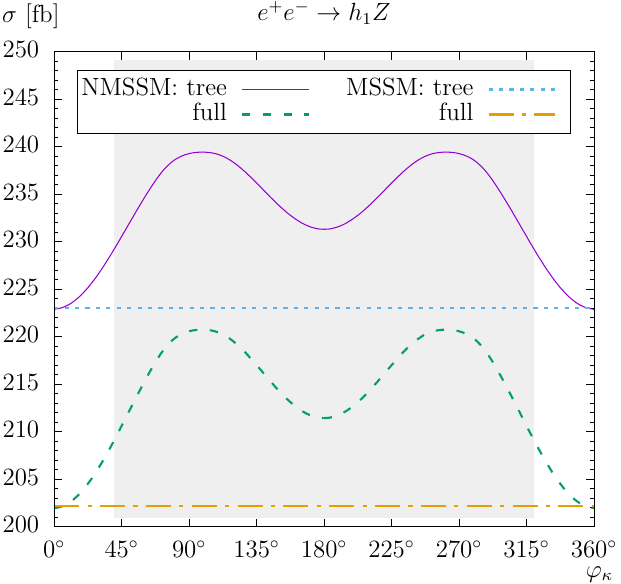}
\includegraphics[width=0.45\textwidth,height=6cm]{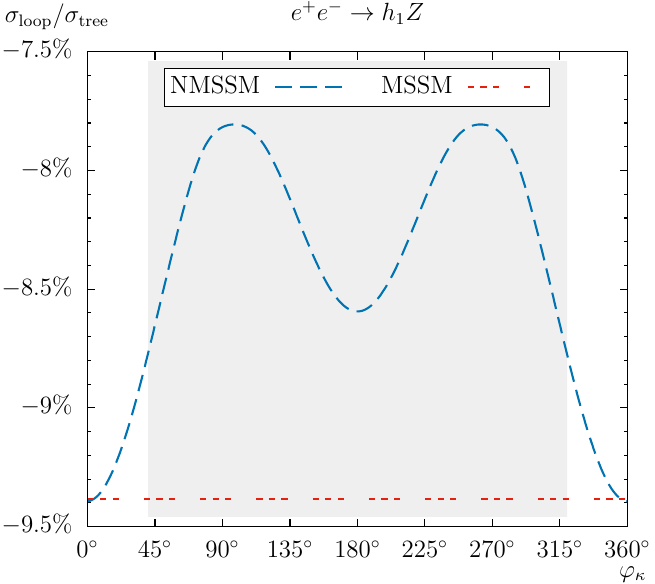}
\\[0.5em]
\includegraphics[width=0.45\textwidth,height=6cm]{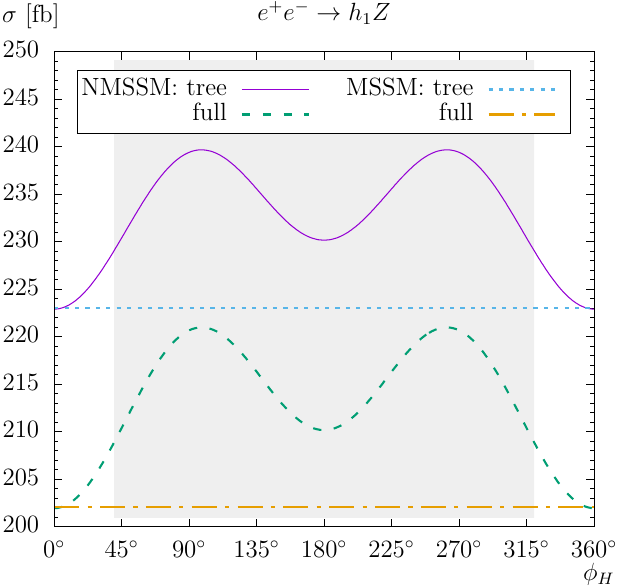}
\includegraphics[width=0.45\textwidth,height=6cm]{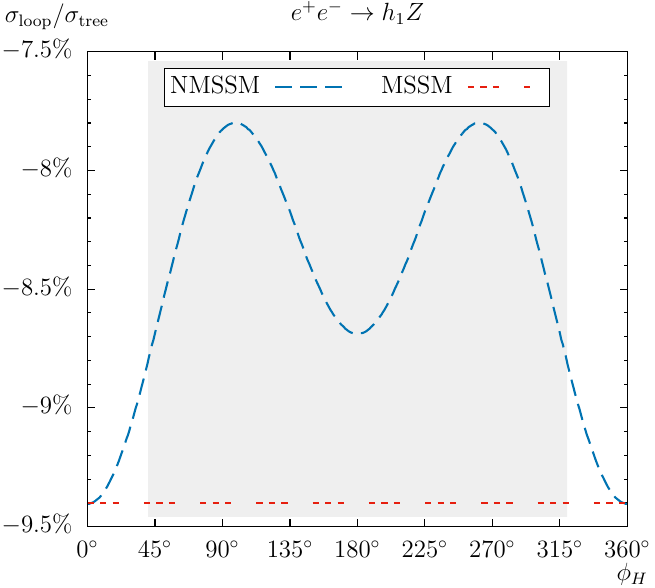}
\end{tabular}
\vspace{-0.5em}
\caption{\label{fig:eeh1Z_5}
  $\sigma(e^+e^- \to h_1 Z)$.
  Tree-level and full one-loop corrected cross sections in the (N)MSSM are
  shown  with parameters chosen according to \refta{tab:para} (\OS).  
  The upper plots show the cross sections with the phase
  $\varphi_{\la}$ varied (left) and the corresponding
  $\sigma_{\mathrm{loop}}/\sigma_{\mathrm{tree}}$ in percent (right);
  the middle plots show the cross sections with the phase
  $\varphi_{\ka}$ varied (left) and the corresponding
  $\sigma_{\mathrm{loop}}/\sigma_{\mathrm{tree}}$ in percent (right);  
  the lower plots show the phase $\phi_H$ varied (left) and the
  corresponding $\sigma_{\mathrm{loop}}/\sigma_{\mathrm{tree}}$ in percent
  (right). The light-gray area is excluded by $\mh1 < 123.5 \gev$.
}
\end{center}
\vspace{-1em}
\end{figure}

In the upper left plot of \reffi{fig:eeh1Z_5} the dependence of the
Higgs-boson production cross section in the (N)MSSM on $\varphi_{\la}$
is depicted. Overall
a relevant variation can be observed. This occurs, however, mostly in
the light gray region with $\mh1 < 123.5 \gev$. As in the previous
section the variation of the cross section is mainly a kinematic effect
due to the ``allowed'' variation of $\mh1$. 
The size of the loop corrections, as before, is found at the level of
about $-9.5\%$, see the upper right plot in \reffi{fig:eeh1Z_5}.
The peaks and dips around $30^{\circ}$, $90^{\circ.}$, $150^{\circ}$,
$210^{\circ}$, $270^{\circ}$, and $330^{\circ}$ in the $\varphi_{\la}$
are due to a numerical instability in \texttt{NMSSMCALC}.

% Fig8
%%%%%%%%%%%%%%%%%%%%%%%%%% F I G U R E %%%%%%%%%%%%%%%%%%%%%%%%%%%%%%%%%%%%%%%
%%%%%%%%%%%%%%%%%%%%%%%%%% F I G U R E %%%%%%%%%%%%%%%%%%%%%%%%%%%%%%%%%%%%%%%

In the middle left plot of \reffi{fig:eeh1Z_5} the (N)MSSM cross sections
are shown in dependence of $\varphi_{\ka}$. As for the dependence on
$\varphi_{\la}$, the visible effects are of pure kinematical nature, as
can be seen in the comparison to the $\varphi_{\ka}$-independent cross
sections in the MSSM. The size of the loop corrections stays at the
level of about $-9.5\%$, see the middle right plot of
\reffi{fig:eeh1Z_5}. The dependence on $\phi_H$ is nearly identical,
as shown in the lower row of \reffi{fig:eeh1Z_5}.

% Fig9
%%%%%%%%%%%%%%%%%%%%%%%%%% F I G U R E %%%%%%%%%%%%%%%%%%%%%%%%%%%%%%%%%%%%%%%
\begin{figure}[ht!]
\begin{center}
\begin{tabular}{c}
\includegraphics[width=0.48\textwidth,height=6cm]{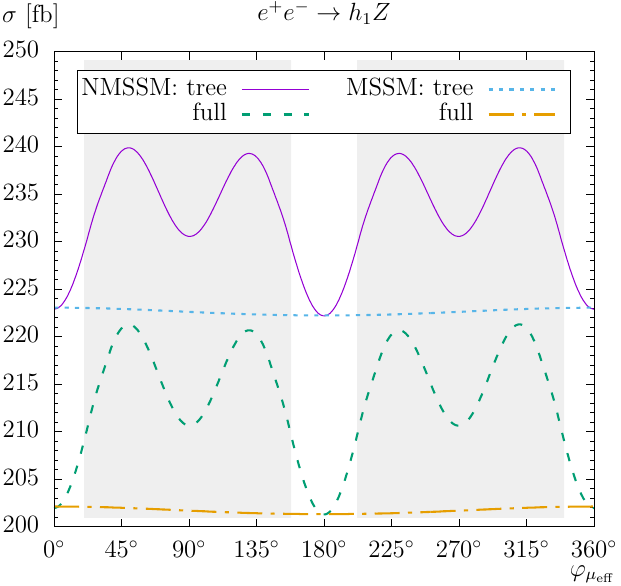}
\includegraphics[width=0.48\textwidth,height=6cm]{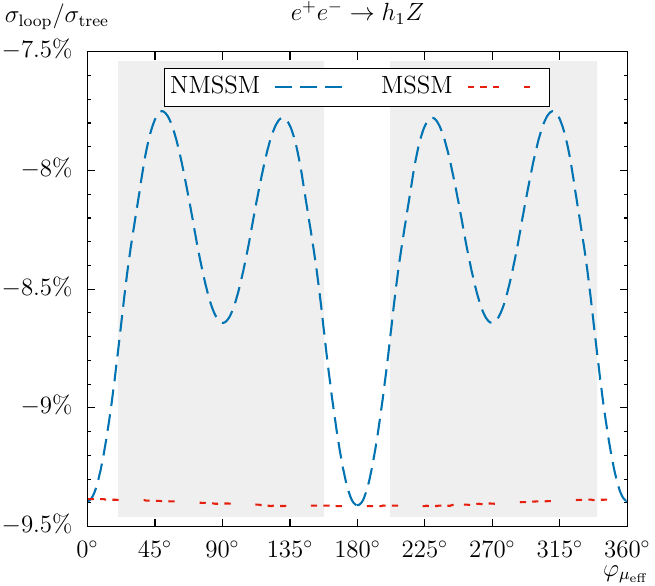}
\\[1em]
\includegraphics[width=0.48\textwidth,height=6cm]{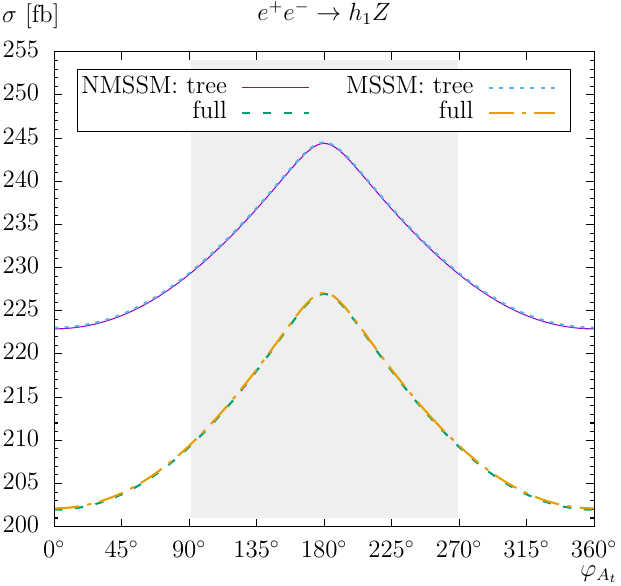}
\includegraphics[width=0.48\textwidth,height=6cm]{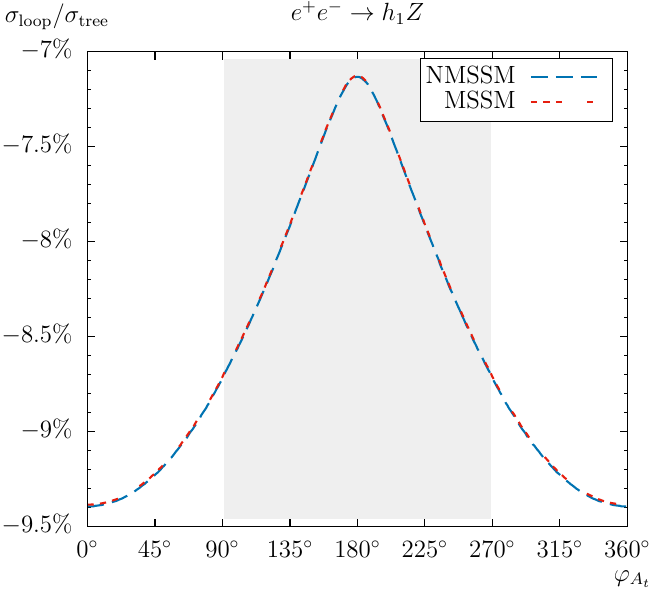}
\end{tabular}
\caption{\label{fig:eeh1Z_6}
  $\sigma(e^+e^- \to h_1 Z)$.
  Tree-level and full one-loop corrected cross sections in the (N)MSSM are
  shown with parameters chosen according to \refta{tab:para} (\OS).  
  The upper plots show the cross sections with the phase
  $\varphi_{\mueff}$ varied (left) and the corresponding
  $\sigma_{\mathrm{loop}}/\sigma_{\mathrm{tree}}$ in percent (right);
  the lower plots show the phase $\varphi_{A_t}$ varied (left) and
  the corresponding $\sigma_{\mathrm{loop}}/\sigma_{\mathrm{tree}}$ in percent
  (right).
  The light-gray area is excluded by $\mh1 \approx M_h < 123.5 \gev$.
}
\end{center}
\end{figure}
%%%%%%%%%%%%%%%%%%%%%%%%%% F I G U R E %%%%%%%%%%%%%%%%%%%%%%%%%%%%%%%%%%%%%%%

\medskip

We turn to the variation of the Higgs-boson production cross
sections with $\varphi_{\mueff}$ and $\varphi_{\At}$. The former
dependence is shown in the upper plots of \reffi{fig:eeh1Z_6}. 
As in the two previous cases, only a variation due to ``allowed''
variations in $\mh1$ can be observed.
The loop corrections stay at the level of about $-9.5\%$, see the upper
right plot of \reffi{fig:eeh1Z_6}. 
The maximum deviation of the relative corrections between the NMSSM
and MSSM reach only $0.6\%$ at these four points. 

\medskip

Finally, the full (N)MSSM cross section for $\varphi_{A_t}$ is presented
in the lower plots of \reffi{fig:eeh1Z_6}: the results for the two
models appear indistinguishable. As for the variation with
$|\At|$, also for the variation of $\varphi_{\At}$ no genuine effect
beyond $\mh1 \approx \Mh$ was observed (see also \reffi{fig:eeh1Z_4}).
As before, the size of the loop corrections is found at the level of
$-9.5 \%$ as can be seen in the lower right plot of
\reffi{fig:eeh1Z_6}.

%%%%%%%%%%%%%%%%%%%%%%%%%%%%%%%%%%%%%%%%%%%%%%%%%%%%%%%%%%%%%%%%%%%%%%%%%%%%%

\subsubsection{Summary}

The process $e^+e^- \to h_1 Z$ is of particular interest for colliders
operating at $\sqrt{s} = 250
\gev$~\cite{ilc-web,LCF,fcc-ee-web,cepc-web,LCreport}, but our results can
be taken over to higher energies as well~\cite{ilc-web,LCreport,LCF,clic-web}. 
The precise prediction of the production cross section is crucial for an
accurate determination of the Higgs-boson couplings to SM fermions and
gauge bosons. In our numerical analysis we have demonstrated that the
one-loop corrections in the (N)MSSM are sizable at the level of $-10\%$,
i.e.\ far larger than any anticipated accuracy of the
experimental coupling determination. On the other hand, the deviation
between the NMSSM and MSSM is tiny, below the level of $0.1\%$, once the
Higgs-boson mass is fixed to its experimental value of about $125 \gev$.

%\newpage

%%%%%%%%%%%%%%%%%%%%%%%%%%%%%%%%%%%%%%%%%%%%%%%%%%%%%%%%%%%%%%%%%%%%%%%%%%%%%%%
%%%%%%%%%%%%%%%%%%%%%%%%%%%%%%%%%%%%%%%%%%%%%%%%%%%%%%%%%%%%%%%%%%%%%%%%%%%%%%%

\section{Conclusions}
\label{sec:conclusions}

We evaluated the light neutral (N)MSSM Higgs boson production mode at
$e^+e^-$ colliders with a two-particle final state, \ie $e^+e^- \to h_1 Z$, 
allowing for complex parameters. 
A precise prediction of this Higgs-boson production cross section is
crucial for the high-precision determination of the Higgs-boson
couplings expected at future $e^+e^-$ colliders. 
In order to yield a sufficient accuracy, at least the one-loop
corrections to the  various Higgs boson production modes have to be
considered.  

Our evaluation of the Higgs-boson production process in association with
a $Z$~boson is based on a full one-loop calculation, also including hard,
soft and collinear QED radiation.  The renormalization is chosen to be
identical as for the various Higgs boson decay calculations; see, e.g.,
\citeres{HiggsDecaySferm,HiggsDecayIno}. This allows a consistent
evaluation of the cross section times branching ratio, as it is
required for a precise determination of the Higgs-boson couplings. 

We first very briefly reviewed the relevant sectors including some
details of the one-loop renormalization procedure of the cNMSSM, which 
are relevant for our calculation.
In most cases we follow \citere{MSSMCT,NMSSMCT}. 
We have discussed the calculation of the one-loop diagrams, the
treatment of UV, IR, and collinear divergences that are canceled by the 
inclusion of (hard, soft, and collinear) QED radiation. 
We have checked our result against the literature, 
and found qualitative agreement where possible.

For the numerical analysis we have chosen a parameter set that yields a
light NMSSM Higgs-boson mass of $\mh1 \approx 125 \gev$ for a large
variation of the relevant parameters. Concerning the MSSM, the results
have been obtained in the MSSM-limit of the NMSSM
($\la \to 0$, $\ka \to 0$ with $\la/\ka$ fixed, and 
$\phi_H \to 0$), the light Higgs-boson mass is accordingly found at the
value of $\Mh \approx 125 \gev$ for the relevant parts of the parameter
space (in our benchmark scenario (see \refta{tab:para}) we always
have $h_1 \sim h$).
In our plots the areas with $\mh1 < 123.5 \gev$ (or $\Mh < 123.5 \gev$)
are indicated (and not taken into account in the discussion).
$h_2 \sim A$, and $h_3 \sim H$. 
In the analysis we investigated the variation of the production
cross sections with the center-of-mass energy $\sqrt{s}$, the genuine
NMSSM parameters $\la$, $\ka$, $A_\ka$, as well as with $\MHp$,
$\mueff$, $M_{\tilde Q_3}$, $M_{\tilde U_3}$ and $A_t$. We have also
analyzed the effects of the phases $\varphi_{\la}$, $\varphi_{\ka}$,
$\varphi_{\mueff}$ and $\varphi_{\At}$. For all these analysis (except
the variation with $\sqrt{s}$) we have set the center-of-mass energy to
the first stage value of most of the possible future $e^+e^-$ colliders,
$\sqrt{s} = 250 \gev$.

The loop corrections in the (N)MSSM are found to be sizable, at the
level of $-10\%$. 
i.e.\ far larger than any anticipated accuracy of the
experimental coupling determination. On the other hand, the observed
variations (mostly in the NMSSM) are very small and (nearly) purely
kinematic, originating from the small variation allowed in $\mh1$ due to 
the theoretical uncertainties, see \citere{Slavich:2020zjv}. However,
since the value of the Higgs-boson mass is determined experimentally to
a high accuracy, these kinematic effects must be considered numerical
artefacts. The only genuine loop effects of the NMSSM w.r.t.\ the MSSM were
observed in the dependence on $M_{\tilde Q_3}$, $M_{\tilde U_3}$ and
$A_t$, but well below the \%-level. This also holds for the variation
with the phases included in our analysis.

The numerical results we have shown are, of course, dependent on the choice 
of the SUSY parameters. Nevertheless, they give an idea of the relevance
of the full one-loop corrections. 
Following our analysis it is evident that the full one-loop corrections
are mandatory for a precise prediction of the light c(N)MSSM Higgs boson
production in association with a $Z$~boson. 
The full one-loop corrections must be taken into account in any precise 
determination of (SUSY) parameters from the production of c(N)MSSM Higgs
bosons at future $e^+e^-$ colliders. 
There are plans to implement the evaluation of the Higgs boson 
production into the public code \FH.

%%%%%%%%%%%%%%%%%%%%%%%%%%%%%%%%%%%%%%%%%%%%%%%%%%%%%%%%%%%%%%%%%%%%%%%%%%%%%%

\subsection*{Acknowledgements}
\begingroup
\small
We thank T.~Hahn for his help implementing \texttt{NMSSMCALC} into \FC. 
The work of S.H. has received financial support from the grant
PID2019-110058GB-C21 funded by MCIN/AEI/10.13039/501100011033 and by
``ERDF A way of making Europe'', and in part by by the grant IFT Centro
de Excelencia Severo Ochoa CEX2020-001007-S funded by
MCIN/AEI/10.13039/501100011033. S.H. also acknowledges support from
Grant PID2022-142545NB-C21 funded by
MCIN/AEI/10.13039/501100011033/FEDER, UE.
\endgroup

%%%%%%%%%%%%%%%%%%%%%%%%%%%%%%%%%%%%%%%%%%%%%%%%%%%%%%%%%%%%%%%%%%%%%%%%%%%%%%%
%%%%%%%%%%%%%%%%%%%%%%%%%%%%%%%%%%%%%%%%%%%%%%%%%%%%%%%%%%%%%%%%%%%%%%%%%%%%%%%

\newcommand\jnl[1]{\textit{\frenchspacing #1}}
\newcommand\vol[1]{\textbf{#1}}

\end{document}